\documentclass[prd,preprint,showpacs,groupedaddress]{revtex4-1}
\usepackage{amsmath}
\usepackage{amsfonts}
\usepackage{amssymb}
\usepackage{geometry}
\usepackage{natbib}
\usepackage[english]{babel}
\usepackage{graphicx}
\usepackage{epstopdf}
\usepackage{subfigure}
\usepackage{caption}
\usepackage{multirow}
\usepackage{indentfirst}
\usepackage{mathrsfs}
\usepackage[colorlinks,linkcolor=blue,anchorcolor=blue,citecolor=blue]{hyperref}

\begin{document}

  \setlength{\parindent}{2em}
  \title{Bardeen black hole surrounded by perfect fluid dark matter}
  \author{He-Xu Zhang} \author{Yuan Chen} \author{Tian-Chi Ma} \author{Peng-Zhang He} \author{Jian-Bo Deng} \email[Jian-Bo Deng: ]{dengjb@lzu.edu.cn}
  \affiliation{Institute of Theoretical Physics $\&$ Research Center of Gravitation, Lanzhou University, Lanzhou 730000, China}

  \begin{abstract}
  We derive an exact solution of the spherically symmetric Bardeen black hole surrounded by perfect fluid dark matter (PFDM). By treating the magnetic charge $g$ and dark matter parameter $\alpha$ as thermodynamic variables, we find that the thermodynamic first law and the corresponding Smarr formula are satisfied. The thermodynamic stability of the black hole is also studied. The result show that, there exists a critical radius $r_{+}^{C}$, where the heat capacity diverges, suggesting that the black hole is thermodynamically stable in the range $0<r_{+}<r_{+}^{C}$. In addition, the critical radius $r_{+}^{C}$ increases with the magnetic charge $g$ and decreases with the dark matter parameter $\alpha$. Applying the Newman-Janis algorithm, we generalize the spherically symmetric solution to the corresponding rotating black hole. With the metric at hand, the horizons and ergospheres are studied. It turns out that for a fixed dark matter parameter $\alpha$, in a certain range, with the increase of the rotation parameter $a$ and magnetic charge $g$, the Cauchy horizon radius increases while the event horizon radius decreases. Finally, we investigate the energy extraction by the Penrose process in rotating Bardeen black hole surrounded by PFDM.
  \end{abstract}

  \maketitle
\section{Introduction}

  The singularity theorems proved by Penrose and Hawking, states that under the main assumptions of the strong energy condition holds and the global hyperbolicity exists, in the frame-work of general relativity, every black hole inevitably contains a singularity \cite{hawking1973large}. At the space-time singularity, the curvatures, densities go to infinity \cite{hawking1973large,hawking2010nature} and the predictive power of physical laws is completely broken down. It is widely believed that the space-time singularities are the reflection of the incompleteness of General Relativity, which can be solved in a quantum theory of gravity. Surprisingly, Bardeen in 1968 \cite{bardeen1968non} obtained a black hole solution without a singularity, which can be interpreted as a gravitationally collapsed magnetic monopole arising in a specific form of nonlinear electrodynamics \cite{ayon2000bardeen}. After that, more regular (non-singular) black holes such as Ay\'on-Beato and Garc\'ia black hole \cite{ayon1998regular}, Hayward black hole \cite{hayward2006formation}, and Berej–Matyjasek–Trynieki–Wornowicz black hole \cite{berej2006regular} were proposed. The spherically symmetric Bardeen black hole is described by the metric
  \begin{equation}
  \mathrm{d}s^{2}=-f\left(r\right)\mathrm{d}t^{2}+f\left(r\right)^{-1}\mathrm{d}r^{2}+r^{2}\mathrm{d}\Omega^{2},\quad f\left(r\right)=1-\frac{2Mr^{2}}{\left(r^{2}+g^{2}\right)^{\frac{3}{2}}},
  \end{equation}
  where $g$ and $M$ are the magnetic charge and mass, respectively. It should be noted that, the metric behaves like the Schwarzschild metric at the large distances ($g/r\ll 1$) and near the origin it behaves like de Sitter geometry, which can be realized from
  \begin{equation}\label{introduction}
  	f\left(r\right)\simeq 1-\frac{2M}{g^{3}}r^{2}, \quad \frac{g}{r}\gg 1.
  \end{equation}
  Eq.~(\ref{introduction}) suggests that the Bardeen black hole avoids singularities by a de Sitter core, i.e. the pressure is negative and thus would prevent a singular end-state of the gravitationally collapsed matter \cite{sakharov1966initial,nam2018non}. It is worth noting that all regular black holes (including Bardeen black holes) violate the weak energy condition, however, the region of violation is always shielded by the Cauchy horizon \cite{bambi2013rotating,neves2014regular,kumar2019shadow,ghosh2020ergosphere}. In fact, this violation of classical energy conditions is a natural consequence of the fact that the singularity-free metric might incorporate some quantum gravity effects \cite{kumar2019shadow}.
\par
  Modern cosmological observations reveal that our current universe contains mainly of 4.9$\%$ baryon matter, 26.8$\%$ dark matter, and 68.3$\%$ dark energy, according to the Standard Model of Cosmology \cite{ade2016planck}. Therefore, it is necessary to consider the black hole solutions surrounded by dark matter or dark energy. In recent years, the black hole surrounded by quintessence dark energy have attracted much attention. For example, Kiselev \cite{kiselev2003quintessence} considered the Schwarzschild black hole surrounded by the quintessential energy and then Toshmatov and Stuchl\'ik \cite{toshmatov2017rotating} extended it to the Kerr-like black hole; the quasinormal modes, thermodynamics and phase transition from Bardeen Black hole surrounded by quintessence was discussed by Saleh and Thomas \cite{saleh2018thermodynamics}; the Hayward black holes surrounded by quintessence have been studied in Ref.~\cite{benavides2020rotating}, etc \cite{ghosh2018lovelock,zhang2006quasinormal,ghosh2016rotating,chen2008hawking,abdujabbarov2017shadow,azreg2013thermodynamical}. On the other hand, as one of the dark matter candidates, the perfect fluid dark matter has been proposed by Kiselev \cite{kiselev2003quintessence,kiselev2003quintessential} and further studied in Ref.~\cite{li2012galactic}, which offers a reasonable explanation for the asymptotically flat rotating velocity in spiral galaxies, see Refs.~\cite{haroon2019shadow,hou2018rotating,xu2018kerr1,xu2018kerr2,rizwan2019distinguishing} for more recent researches. In this work, following Refs.~\cite{toshmatov2017rotating,benavides2020rotating}, we generalize the Schwarzschild black hole surrounded by PFDM to the spherically symmetric Bardeen black hole. Furthermore, by resorting to the Newman-Janis algorithm we obtain the rotating Bardeen black hole in PFDM.
\par
  The paper is organized as follows. The next section is the derivation of the spherically symmetric Bardeen black hole surrounded by perfect fluid dark matter. In Sec.~\ref{3}, we discuss its thermodynamic properties. In Sec.~\ref{4}, by applying the Newman-Janis algorithm we obtain the rotating Bardeen black hole surrounded by perfect fluid dark matter. The weak energy condition is the subject of Sec.~\ref{5}. In Sec.~\ref{6}, the horizons and the ergospheres of Bardeen black hole surrounded by perfect fluid dark matter have been studied. In Sec.~\ref{7}, we investigate the energy extraction by the Penrose process. Conclusions and discussions are presented in Sec.~\ref{8}. Planck units $\hbar=G=c=k_{B}=1$ are used throughout the paper.

\section{Static and spherically symmetric Bardeen black hole in perfect fluid dark matter}\label{II}
  Given the coupling between gravitational and a non-linear electromagnetic field, the Einstein-Maxwell equations should be modified as
  \begin{gather}
  	G_{\mu}^{\,\nu}=2\left(\frac{\partial\mathcal{L}\left(F\right)}{\partial F}F_{\mu\lambda}F^{\nu\lambda}-\delta_{\mu}^{\,\nu}\mathcal{L}\right)+8\pi T_{\mu}^{\,\nu},\label{eq:EM1}\\
  	\nabla_{\mu}\left(\frac{\partial\mathcal{L}\left(F\right)}{\partial F}F^{\nu\mu}\right)=0,\label{eq:EM2}\\
  	\nabla_{\mu}\left(*F^{\nu\mu}\right)=0.\label{eq:EM3}
  \end{gather}
  Here, $F_{\mu\nu}=2\nabla_{[\mu}A_{\nu]}$ and $\mathcal{L}$ is a function of $F\equiv\frac{1}{4}F_{\mu\nu}F^{\mu\nu}$ given by \cite{ayon2000bardeen}
  \begin{equation}
  \mathcal{L}\left(F\right)=\frac{3M}{\vert g\vert^{3}}\left(\frac{\sqrt{2g^{2}F}}{1+\sqrt{2g^{2}F}}\right)^{\frac{5}{2}},
  \end{equation}
  where $g$ and $M$ are the parameters associated with magnetic charge and mass, respectively.
\par
  In this work, we consider the black holes surrounded by the perfect fluid dark matter. Following Kiselev \cite{kiselev2003quintessence,kiselev2003quintessential} and Li and Yang \cite{li2012galactic}, the energy-momentum tensor of PFDM in the standard orthogonal basis is given by
  $T_{\nu}^{\,\mu}=\mathrm{diag}(-\epsilon,p_r,p_\theta,p_\phi)$,
  with the density, radial and tangential pressures of the PFDM
  \begin{equation}
  -\epsilon=p_r=\frac{\alpha}{8\pi r^3}\qquad \text{and} \qquad p_\theta=p_\phi=-\frac{\alpha}{16\pi r^3}\ .
  \end{equation}
\par
  To obtain a solution satisfies Eqs.~(\ref{eq:EM1})-(\ref{eq:EM3}), we assume a spherically symmetric line element
  \begin{equation}\label{eq:spherically metric}
   \mathrm{d}s^{2}=-f\left(r\right)\mathrm{d}t^{2}+f\left(r\right)^{-1}\mathrm{d}r^{2}+r^{2}\mathrm{d}\Omega^{2},\quad f\left(r\right)=1-\frac{2m\left(r\right)}{r},
  \end{equation}
  and use the ansatz for Maxwell field \cite{ayon2000bardeen}
  \begin{equation}\label{eq:Maxwell field}
  	F_{\mu\nu}=\left(\delta_{\mu}^{\theta}\delta_{\nu}^{\varphi}-\delta_{\nu}^{\theta}\delta_{\mu}^{\varphi}\right)B\left(r,\theta\right).
  \end{equation}
  Next, using Eqs.~(\ref{eq:EM2}) and (\ref{eq:EM3}), Eq.~(\ref{eq:Maxwell field}) can be simplified as
  \begin{equation}
   F_{\mu\nu}=\left(\delta_{\mu}^{\theta}\delta_{\nu}^{\varphi}-\delta_{\nu}^{\theta}\delta_{\mu}^{\varphi}\right)g\sin{\theta},
  \end{equation}
  where $g$ is the integration constant. Further, one can get $F=g^{2}/2r^{4}$.
  In order to give a direct physical interpretation to $g$, for any 2-sphere $S$ at infinity, we consider the following integral
  \begin{equation}\label{eq:g}
  \frac{1}{4\pi}\int_{S}*F=\frac{g}{4\pi}\int_{0}^{\pi}\int_{0}^{2\pi}\sin{\theta}\mathrm{d}\theta\mathrm{d}\varphi=g.
  \end{equation}
  From Eq.~(\ref{eq:g}), one can confirm that $g$ is the magnetic monopole charge.
\par
  Now, with the help of the above equations, the time component of Eq.~(\ref{eq:EM1}) reduces to
  \begin{equation}\label{eq:integral}
  -\frac{2}{r^{2}}\frac{\mathrm{d}m\left(r\right)}{\mathrm{d}r}=-\frac{6Mg^{2}}{\left(r^{2}+g^{2}\right)^{\frac{5}{2}}}+\frac{\alpha}{r^{3}}.
  \end{equation}
  Integrating Eq.~(\ref{eq:integral}) from $r$ to $\infty$ and using that $M=\lim_{r\to\infty}\left(m\left(r\right)+\frac{\alpha}{2}\ln{\frac{r}{\vert\alpha\vert}}\right)$, one finally gets
  \begin{equation}\label{eq:f}
  f\left(r\right)=1-\frac{2Mr^{2}}{\left(r^{2}+g^{2}\right)^{\frac{3}{2}}}+\frac{\alpha}{r}\ln{\frac{r}{\vert\alpha\vert}}.
  \end{equation}
  Note that, in the absence of PFDM, i.e. $\alpha=0$, the above space-time recovers that of the Bardeen black hole; in the case of $g=0$ \cite{bardeen1968non,ayon2000bardeen}, it reduces to the Schwarzschild black hole surrounded by PFDM \cite{li2012galactic,kiselev2003quintessential}; if $\alpha=0$ and $g=0$, we will obtain the Schwarzschild black hole.
\par
   As was mentioned in the introduction, the Bardeen black hole ($\alpha=0$) is regular everywhere. To check whether this character is changed by the presence of perfect fluid dark matter, we calculate the following curvature scalars
  \begin{align}
  R&=\frac{6M g^{2}\left(4g^{2}-r^{2}\right)}{\left(g^{2}+r^{2}\right)^{\frac{7}{2}}}-\frac{\alpha}{r^{3}},\\
  R_{\mu\nu}&R^{\mu\nu}=\frac{18M^{2}g^{4}\left(8g^{4}-4g^{2}r^{2}+13r^{4}\right)}{\left(g^{2}+r^{2}\right)^{7}}-\frac{6Mg^{2}\left(2g^{2}+7r^{2}\right)\alpha}{r^{3}\left(g^{2}+r^{2}\right)^{\frac{7}{2}}}+\frac{5\alpha^{2}}{2r^{6}},\\
  \mathcal{K}=&R_{\mu\nu\sigma\tau}R^{\mu\nu\sigma\tau}=\frac{12 \alpha^2 \ln^2\frac{r}{\left| \alpha\right| }}{r^6}+\frac{13 \alpha^2}{r^6}+\frac{4 \alpha \ln\frac{r}{\left|\alpha\right| } \left[\frac{6 M r^5 \left(3 g^2-2 r^2\right)}{\left(g^2+r^2\right)^{7/2}}-5 \alpha\right]}{r^6}\\\nonumber
  +&\frac{4 \alpha M \left(-2g^4-37 g^2 r^2+10 r^4\right)}{r^3 \left(g^2+r^2\right)^{7/2}}+\frac{12 M^2 \left(8 g^8-4 g^6 r^2+47 g^4 r^4-12 g^2 r^6+4 r^8\right)}{\left(g^2+r^2\right)^7}.
  \end{align}
  It turns out that the Bardeen black hole surrounded by PFDM is singular at $r=0$; in fact, the future singularity comes from the dark matter background.

\section{Thermodynamic properties of Bardeen black hole in PFDM}{\label{3}}
Let us now turn to the thermodynamic properties of the Bardeen black hole in perfect fluid dark matter. For convenience, we write here the line element of spherically symmetric balck hole obtained in the previous section as
\begin{equation}\label{static line element}
\begin{gathered}
\mathrm{d}s^{2}=-f\left(r\right)\mathrm{d}t^{2}+f\left(r\right)^{-1}\mathrm{d}r^{2}+r^{2}\mathrm{d}\Omega^{2},\\ f\left(r\right)=1-\frac{2Mr^{2}}{\left(r^{2}+g^{2}\right)^{\frac{3}{2}}}+\frac{\alpha}{r}\ln{\frac{r}{\vert\alpha\vert}}.
\end{gathered}
\end{equation}
\par
The black hole mass $M$ can be expressed in terms of event horizon $r_+$ as
\begin{equation}\label{black hole mass}
      M=\frac{\left(r_{+}^{2}+g^{2}\right)^{\frac{3}{2}}}{2r_{+}^{2}}\left(1+\frac{\alpha}{r_{+}}\ln\frac{r_{+}}{\vert\alpha\vert}\right),
\end{equation}
which comes from $f\left(r_{+}\right)=0$. According to the Bekenstein area law, the entropy $S$ of black hole can be calculated as
\begin{equation}\label{S}
S=\frac{\mathcal{A}}{4}=\int_{0}^{2\pi}\int_{0}^{\pi}\sqrt{g_{\theta\theta}g_{\varphi\varphi}}\mathrm{d}\theta\mathrm{d}\varphi=\pi r_{+}^{2}.
\end{equation}
Making use of it, one can rewrite Eq.~(\ref{black hole mass}) as
\begin{equation}\label{black hole mass2}
M=\frac{\left(S+\pi g^{2}\right)^{\frac{3}{2}}}{2\sqrt{\pi}S}\left(1+\frac{\sqrt{\pi}\alpha}{\sqrt{S}}\ln\frac{\sqrt{S}}{\sqrt{\pi}\vert\alpha\vert}\right).
\end{equation}
Other thermodynamic quantities can be obtained through thermodynamic identities. For examples, the temperature $T$, the magnetic potential $\Psi$ and the conjugate quantity to the dark matter parameter $\alpha$ are given by
\begin{align}
   T&=\left(\frac{\partial M}{\partial S}\right)_{g,\alpha}=\frac{\sqrt{r_{+}^2+g^2}\left(r_+\left(r_{+}^{2}-2g^{2}\right)+\alpha\left(r_{+}^{2}+g^{2}\right)-3g^{2}\alpha\ln{\frac{r_{+}}{\vert\alpha\vert}}\right)}{4\pi r_{+}^{5}},\label{T}\\
   \Psi&=\left(\frac{\partial M}{\partial g}\right)_{T,\alpha}=\frac{3g\sqrt{r_{+}^{2}+g^2}\left(r_++\alpha\ln\frac{r_{+}}{\vert\alpha\vert}\right)}{2r_{+}^3},\\
   \Pi&=\left(\frac{\partial M}{\partial \alpha}\right)_{T,g}=\frac{(r_{+}^{2}+g^2)^{\frac{3}{2}}\left(-1+\ln\frac{r_{+}}{\vert\alpha\vert}\right)}{2r_{+}^3}.
\end{align}
Note that here we have treated the dark matter parameter $\alpha$ as a new thermodynamic variable and $\Pi$ is its conjugate quantity, as shown in Refs.~\cite{balart2017smarr,cai2013pv,xu2019perfect,wei2020extended}. It is easy to check that those thermodynamic quantities satisfy the first law of black hole thermodynamics
\begin{equation}
   \mathrm{d}M=T\mathrm{d}S+\Psi\mathrm{d}g+\Pi\mathrm{d}\alpha,
\end{equation}
and the Smarr formula
\begin{equation}
   M=2TS+\Psi g+\Pi \alpha,
\end{equation}
which is exactly consistent with the scaling dimensional argument.
\par
\begin{figure}[htbp]
	\centering
	\includegraphics[width=.49\textwidth]{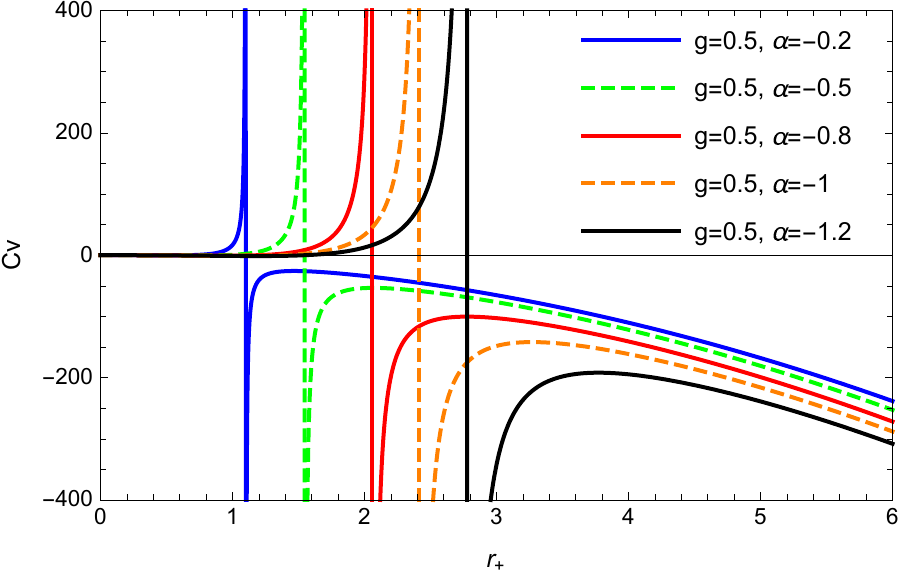}
	\includegraphics[width=.49\textwidth]{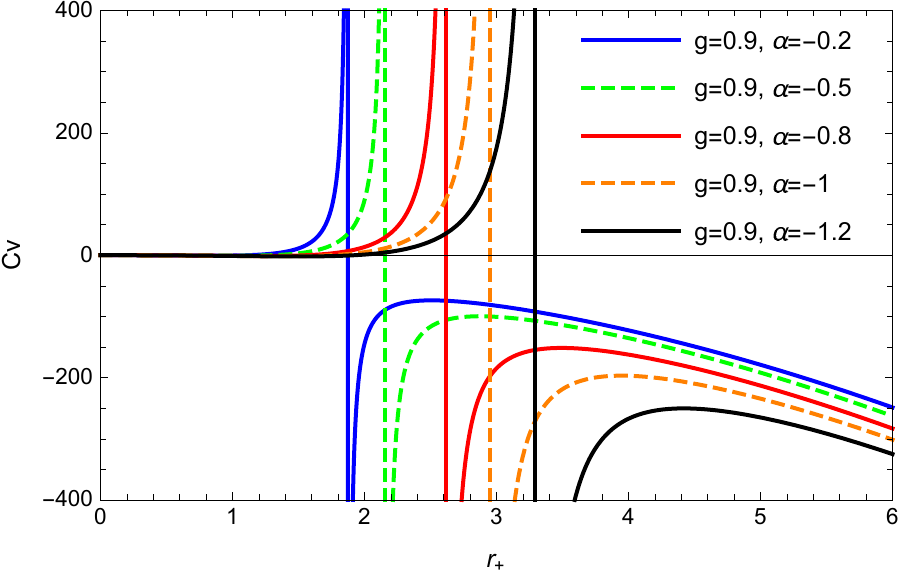}
	\includegraphics[width=.49\textwidth]{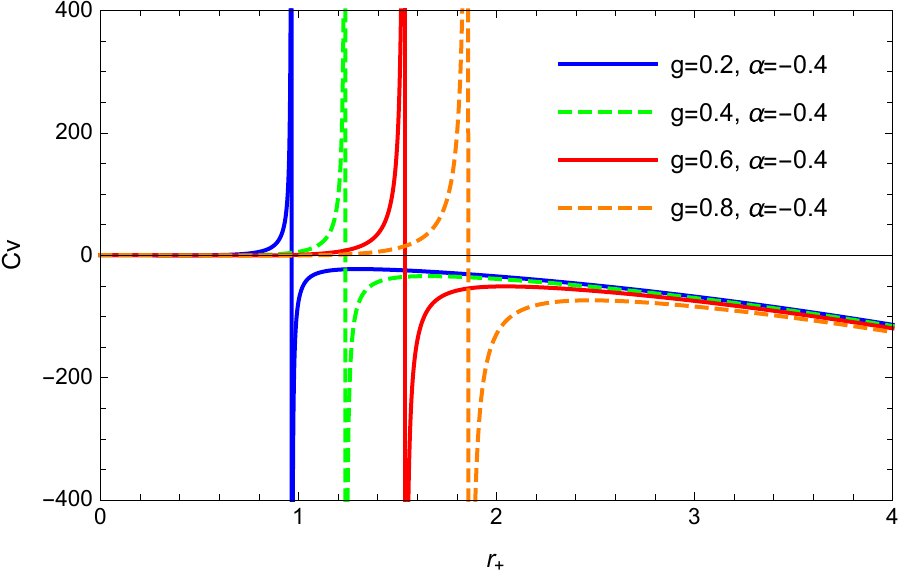}
	\includegraphics[width=.49\textwidth]{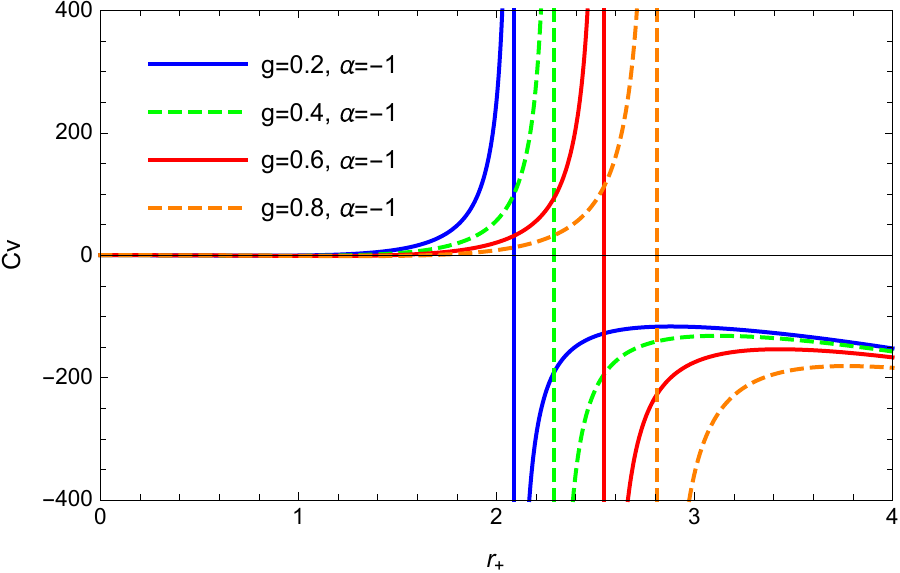}
	\caption{Variation of heat capacity with respect to the event horizon $r_{+}$ for a set of values of parameters $g$ and $\alpha$.}
	\label{heat c}
\end{figure}

In what follows, we will investigate the thermodynamic stability of the Bardeen black hole in perfect fluid dark matter. The heat capacity at constant volume is defined as
\begin{equation}\label{heat capacity}
C_{V}=T\frac{\partial{S}}{\partial{T}}=T\frac{\partial{S}}{\partial{r_{+}}}\left(\frac{\partial{T}}{\partial{r_{+}}}\right)^{-1}.
\end{equation}
Plugging Eqs. (\ref{S}) and (\ref{T}) into (\ref{heat capacity}), one finally arrives at
\begin{equation}
C_{V}=-\frac{2\pi r_{+}^{2}\left(r_{+}^{2}+g^{2}\right)\left(r_{+}^{3}+\alpha r_{+}^{2}-2g^{2}r_{+}+g^{2}\alpha-3g^{2}\alpha\ln\frac{r_{+}}{\vert\alpha\vert}\right)}{r_{+}^{4}\left(r_++2\alpha\right)-8g^{4}\left(r_+-\alpha\right)+2g^{2}r_{+}^{2}\left(5\alpha-2r_{+}\right)-3g^{2}\left(5g^{2}+4r_{+}^{2}\right)\alpha\ln\frac{r_{+}}{\vert\alpha\vert}}.
\end{equation}
The behavior of $C_{V}$  against $r_{+}$ was plotted in Fig.~\ref{heat c} for different values of the magnetic charge $g$ and the dark matter parameter $\alpha$. As we can see, for given values of parameters $a$ and $g$, there is a critical radius $r_{+}^{C}$ where heat capacity $C_{V}$ diverge and the second order phase transition occurs. The Fig.~\ref{heat c} indicates the heat capacity is positive and the black hole is thermodynamically stable in the range $0<r_{+}<r_{+}^{C}$. Clearly, one can find that the critical radius $r_{+}^{C}$ increases with the magnetic charge $g$ and decreases with the dark matter parameter $\alpha$.

\section{Rotating Bardeen black hole in perfect fluid dark matter}\label{4}
  In this section, with the Newman-Janis algorithm (NJA), we will generalize the spherically symmetric Bardeen black hole solution in perfect fluid dark matter to Kerr-like rotational black hole solution. The NJA was first proposed by Newman and Janis in 1965 \cite{newman1965note} and widely used in many articles \cite{toshmatov2017generic,benavides2020rotating,toshmatov2017rotating,kumar2018rotating,xu2017kerr,xu2020black,shaikh2019black,kim2020rotating,jusufi2020rotating,azreg2019rotating,liu2020shadow}. In this work, we will adopt the NJA modified by Azreg-A\"inou \cite{azreg2014generating,azreg2014static}, which can generate rotating regular black hole solutions without complexification.
\par
  Consider the general static and spherically symmetric metric:
  \begin{equation}\label{eq:general sphere}
  \mathrm{d}s^{2}=-f\left(r\right)\mathrm{d}t^{2}+g\left(r\right)^{-1}\mathrm{d}r^{2}+h\left(r\right)\mathrm{d}\Omega^{2},\quad \mathrm{d}\Omega^{2}=\mathrm{d}\theta^{2}+\sin^{2}{\theta}\mathrm{d}\varphi^{2}.
  \end{equation}
  At the first step of this algorithm, we transform the spherically symmetric space-time metric (\ref{eq:general sphere}) from the Boyer-Lindquist (BL) coordinates ($t, r, \theta, \varphi$) to the Eddington-Finkelstein (EF) coordinates ($u, r, \theta, \varphi$) by carrying out the coordinate transformation
  \begin{equation}
  \mathrm{d}u=\mathrm{d}t-\frac{\mathrm{d}r}{\sqrt{f(r)g(r)}}.
  \end{equation}
  As the result of this transformation, the line element (\ref{eq:general sphere}) takes the form
  \begin{equation}\label{eq:EF}
  \mathrm{d}s^{2}=-f\left(r\right)\mathrm{d}u^{2}-2\sqrt{\frac{f\left(r\right)}{g\left(r\right)}}\mathrm{d}u\mathrm{d}r+h\left(r\right)\left(\mathrm{d}\theta^{2}+\sin^{2}{\theta}\mathrm{d}\varphi^{2}\right).
  \end{equation}
  In terms of the null tetrads satisfy the relations $l_{\mu}l^{\mu}=n_{\mu}n^{\mu}=m_{\mu}m^{\mu}=l_{\mu}m^{\mu}=n_{\mu}m^{\mu}=0, l_{\mu}n^{\mu}=-m_{\mu}\bar{m}^{\mu}=1$, the contravariant metric tensor associated with the line element (\ref{eq:EF}) can be expressed as
  \begin{equation}\label{eq:covmetric}
  g^{\mu\nu}=-l^{\mu}n^{\nu}-l^{\nu}n^{\mu}+m^{\mu}\bar{m}^{\nu}+m^{\nu}\bar{m}^{\mu},
  \end{equation}
  where
  \begin{equation}
  \begin{gathered}
  l^{\mu}=\delta^{\mu}_{r},\qquad n^{\mu}=\sqrt{\frac{g\left(r\right)}{f\left(r\right)}}\delta^{\mu}_{0}-\frac{f\left(r\right)}{2}\delta^{\mu}_{r},\\
  m^{\mu}=\frac{1}{\sqrt{2h\left(r\right)}}\delta^{\mu}_{\theta}+\frac{i}{\sqrt{2h\left(r\right)}\sin{\theta}}\delta^{\mu}_{\varphi},\quad
  \bar{m}^{\mu}=\frac{1}{\sqrt{2h\left(r\right)}}\delta^{\mu}_{\theta}-\frac{i}{\sqrt{2h\left(r\right)}\sin{\theta}}\delta^{\mu}_{\varphi}.
  \end{gathered}
  \end{equation}
  Next, we take the critical step of the NJA, which is to perform complex coordinate transformations in the $u-r$ plane
  \begin{equation}
  \begin{gathered}
  u\rightarrow u-ia\cos{\theta},\\
  r\rightarrow r+ia\cos{\theta}.
  \end{gathered}
  \end{equation}
  At the same time, we assume that as the result of these transformations the metric functions also turn into a new form: $f\left(r\right)\rightarrow F\left(r, a, \theta\right)$, $g\left(r\right)\rightarrow G\left(r, a, \theta\right)$, and $h\left(r\right)\rightarrow\Sigma=r^{2}+a^{2}\cos^{2}{\theta}$ \cite{azreg2014generating,azreg2014static}. Furthermore, null tetrads also take the new form
  \begin{equation}
  \begin{split}
  l^{\mu}&=\delta^{\mu}_{r},\quad n^{\mu}=\sqrt{\frac{G}{F}}\delta^{\mu}_{0}-\frac{1}{2}F\delta^{\mu}_{r},\\
  m^{\mu}&=\frac{1}{\sqrt{2\Sigma}}\left(\delta^{\mu}_{\theta}+ia\sin{\theta}\left(\delta^{\mu}_{0}-\delta^{\mu}_{r}\right)+\frac{i}{\sin{\theta}}\delta^{\mu}_{\varphi}\right),\\
  \bar{m}^{\mu}&=\frac{1}{\sqrt{2\Sigma}}\left(\delta^{\mu}_{\theta}-ia\sin{\theta}\left(\delta^{\mu}_{0}-\delta^{\mu}_{r}\right)-\frac{i}{\sin{\theta}}\delta^{\mu}_{\varphi}\right).
  \end{split}
  \end{equation}
  Then by means of Eq. (\ref{eq:covmetric}), the contravariant components of the metric $g^{\mu\nu}$ can be obtained as
  \begin{equation}
  \begin{split}
  g^{uu}&=\frac{a^{2}\sin^{2}{\theta}}{\Sigma},\quad g^{rr}=G+\frac{a^{2}\sin^{2}{\theta}}{\Sigma},\\
  g^{\theta\theta}&=\frac{1}{\Sigma},\quad g^{\varphi\varphi}=\frac{1}{\Sigma\sin^{2}{\theta}},\\
  g^{ur}&=g^{ru}=-\sqrt{\frac{G}{F}}-\frac{a^{2}\sin^{2}{\theta}}{\Sigma},\\
  g^{u\varphi}&=g^{\varphi u}=\frac{a}{\Sigma},\quad g^{r\varphi}=g^{\varphi r}=-\frac{a}{\Sigma}.
  \end{split}
  \end{equation}
  Accordingly, the rotating metric in the EF coordinates of ($u, r, \theta, \varphi$) reads
  \begin{equation}\label{eq:rotating EF}
  \begin{split}
  \mathrm{d}s^{2}=&-F\mathrm{d}u^{2}-2\sqrt{\frac{F}{G}}\mathrm{d}u\mathrm{d}r+2a\left(F-\sqrt{\frac{F}{G}}\right)\sin^{2}{\theta}\mathrm{d}u\mathrm{d}\varphi+\Sigma\mathrm{d}\theta^{2}\\
  &+2a\sin^{2}{\theta}\sqrt{\frac{F}{G}}\mathrm{d}r\mathrm{d}\varphi+\sin^{2}{\theta}\left[\Sigma+a^{2}\left(2\sqrt{\frac{F}{G}}-F\right)\sin^{2}{\theta}\right]\mathrm{d}\varphi^{2}.
  \end{split}
  \end{equation}
\par
  The last step of NJA is to bring (\ref{eq:rotating EF}) to the BL coordinates by using the following
  coordinate transformations:
  \begin{equation}
  \mathrm{d}u=\mathrm{d}t+\lambda\left(r\right)\mathrm{d}r,\quad \mathrm{d}\varphi=\mathrm{d}\phi+\chi\left(r\right)\mathrm{d}r,
  \end{equation}
  where the functions $\lambda\left(r\right)$ and $\chi\left(r\right)$ can be found using the requirement that all the nondiagonal components of the metric tensor, except the coefficient $g_{t\phi}$ $(g_{\phi t})$, are equal to zero~\cite{azreg2014generating,azreg2014static}. Thus,
  \begin{equation}
  \lambda\left(r\right)=-\frac{k(r)+a^{2}}{g\left(r\right)h\left(r\right)+a^{2}},\quad \chi\left(r\right)=-\frac{a}{g\left(r\right)h\left(r\right)+a^{2}},
  \end{equation}
  with
  \begin{equation}
  k\left(r\right)=\sqrt{\frac{g\left(r\right)}{f\left(r\right)}}h\left(r\right),
  \end{equation}
  and
  \begin{equation}
   F\left(r,\theta\right)=\frac{\left(gh+a^{2}\cos^{2}{\theta}\right)\Sigma}{\left(k^{2}+a^{2}\cos^{2}{\theta}\right)^{2}}, \quad G\left(r,\theta\right)=\frac{gh+a^{2}\cos^{2}{\theta}}{\Sigma}.
  \end{equation}
  Here, for convenience, we omit the variables of $f(r)$, $g(r)$, $h(r)$ and $k(r)$.\\
  Finally, the rotating solution corresponding to the spherically symmetric metric (\ref{eq:general sphere}) can therefore  be obtained as
  \begin{equation}\label{eq:rotating Kerr}
  \begin{split}
  \mathrm{d}s^{2}=&-\frac{\left(gh+a^{2}\cos^{2}{\theta}\right)\Sigma}{\left(k+a^{2}\cos^{2}{\theta}\right)^{2}}\mathrm{d}t^{2}+\frac{\Sigma}{gh+a^{2}}\mathrm{d}r^{2}-2a\sin^{2}\theta\left[\frac{k-gh}{\left(k+a^{2}\cos^{2}{\theta}\right)^{2}}\right]\Sigma\mathrm{d}\phi\mathrm{d}t\\
  &+\Sigma\mathrm{d}\theta^{2}+\Sigma\sin^{2}\theta\left[1+a^{2}\sin^{2}\theta\frac{2k-gh+a^{2}\cos^{2}\theta}{\left(k+a^{2}\cos^{2}{\theta}\right)^{2}}\right]\mathrm{d}\phi^{2}.
  \end{split}
  \end{equation}
\par
  In the case of Bardeen black holes in PFDM, comparing the line elements (\ref{static line element}) with (\ref{eq:general sphere}), one can find
  \begin{equation}
  f\left(r\right)=g\left(r\right)=1-\frac{2Mr^{2}}{\left(r^{2}+g^{2}\right)^{\frac{3}{2}}}+\frac{\alpha}{r}\ln{\frac{r}{\vert\alpha\vert}}, \quad h\left(r\right)=k\left(r\right)=r^{2}.
  \end{equation}
\par
  Substituting the above expressions into (\ref{eq:rotating Kerr}), we therefore obtain the metric of rotating Bardeen black holes in perfect fluid dark matter in the form
  \begin{equation}\label{eq:rotating metric}
  \begin{split}
  \mathrm{d}s^{2}=&-\left(1-\frac{2\rho r}{\Sigma}\right)\mathrm{d}t^{2}+\frac{\Sigma}{\Delta_{r}}\mathrm{d}r^{2}+\Sigma\mathrm{d}\theta^{2}-\frac{4a\rho r\sin^{2}{\theta}}{\Sigma}\mathrm{d}t\mathrm{d}\phi\\&+\sin^{2}{\theta}\left(r^{2}+a^{2}+\frac{2a^{2}\rho r\sin^{2}{\theta}}{\Sigma}\right)\mathrm{d}\phi^{2},
  \end{split}
  \end{equation}
  with
  \begin{equation}\label{eq:line element terms}
  \begin{gathered}
  2\rho=\frac{2Mr^{3}}{\left(r^{2}+g^{2}\right)^{\frac{3}{2}}}-\alpha\ln{\frac{r}{\vert\alpha\vert}},\\
  \Sigma=r^{2}+a^{2}\cos^{2}{\theta},\\
  \Delta_{r}=r^{2}+a^{2}-\frac{2Mr^{4}}{\left(r^{2}+g^{2}\right)^{\frac{3}{2}}}+\alpha r\ln{\frac{r}{\vert\alpha\vert}}.
  \end{gathered}
  \end{equation}
\par
  Now, we come to consider the energy-momentum tensor in the following. With the help of $Mathematica$ package, from the metric (\ref{eq:rotating metric}), the nonvanishing components of the Einstein tensor $G_{\mu\nu}$ are given by
  \begin{equation}\label{eq:Einstein tensor}
  \begin{aligned}
  G_{tt}=&\frac{2\left(r^{4}+a^{2}r^{2}-2r^{3}\rho-a^{4}\sin^{2}{\theta}\cos^{2}{\theta}\right)\rho'}{\Sigma^{3}}-\frac{a^{2}r\sin^{2}{\theta}\rho''}{\Sigma^{2}},\\
  G_{t\phi}=&\frac{2a\sin^{2}{\theta}\left[\left(r^{2}+a^{2}\right)\left(a^{2}\cos^{2}{\theta}-r^{2}\right)+2r^{3}\rho\right]\rho'}{\Sigma^{3}}\\
  &+\frac{ar\sin^{2}{\theta}\left(r^{2}+a^{2}\right)\rho''}{\Sigma^{2}},\\
  G_{rr}=&-\frac{2r^{2}\rho'}{\Sigma\Delta_{r}},\quad G_{\theta\theta}=-\frac{2a^{2}\cos^{2}{\theta}\rho'}{\Sigma}-r\rho'',\\
  G_{\phi\phi}=&-\frac{a^{2}\sin^{2}{\theta}\left(r^{2}+a^{2}\right)\left[a^{2}+\left(2r^{2}+a^{2}\right)\cos{2\theta}\right]}{\Sigma^{3}}-\frac{4a^{2}r^{3}\sin^{4}{\theta}\rho\rho'}{\Sigma^{3}}-\frac{r\sin^{2}{\theta}\left(a^{2}+r^{2}\right)^{2}\rho''}{\Sigma^{2}},
  \end{aligned}
  \end{equation}
  in which the prime $'$ denotes the derivative with respect to $r$.
\par
  In order to obtain the components of the energy-momentum tensor, as shown in Refs.~\cite{toshmatov2017rotating,benavides2020rotating}, we introduce the standard orthonormal basis of the rotating Bardeen black hole in perfect fluid dark matter
  \begin{equation}\label{eq:basis}
  \begin{split}
  e^{\mu}_{\left(t\right)}=&\frac{1}{\sqrt{\Delta_{r}\Sigma}}\left(r^{2}+a^{2},0,0,a\right),\\
  e^{\mu}_{\left(r\right)}=&\sqrt{\frac{\Delta_{r}}{\Sigma}}\left(0,1,0,0\right),\\
  e^{\mu}_{\left(\theta\right)}=&\frac{1}{\sqrt{\Sigma}}\left(0,0,1,0\right),\\
  e^{\mu}_{\left(\phi\right)}=&-\frac{1}{\sqrt{\Sigma\sin^{2}{\theta}}}\left(a\sin^{2}{\theta},0,0,1\right).
  \end{split}
  \end{equation}
  Combining (\ref{eq:Einstein tensor}), (\ref{eq:basis}), and the Einstein field equation $G_{\mu\nu}=8\pi T_{\mu\nu}$, the components of the energy-momentum tensor can be obtained as
  \begin{equation}\label{eq:state equations}
  \begin{split}
  \epsilon&=-p_{r}=T_{\left(t\right)\left(t\right)}=\frac{1}{8\pi}e^{\mu}_{\left(t\right)}e^{\nu}_{\left(t\right)}G_{\mu\nu}=\frac{1}{4\pi}\frac{r^{2}\rho'}{\Sigma^{2}},\\
  p_{\theta}&=p_{\phi}=T_{\left(\theta\right)\left(\theta\right)}=\frac{1}{8\pi}e^{\mu}_{\left(\theta\right)}e^{\nu}_{\left(\theta\right)}G_{\mu\nu}=\epsilon-\frac{2\rho'+r\rho''}{8\pi\Sigma}.
  \end{split}
  \end{equation}
  From Eq. (\ref{eq:state equations}), it is easy to find that $\epsilon$, $p_{r}$, $p_{\theta}$, and $p_{\phi}$ all contain two contributions: a nonlinear magnetic-charged part and the PFDM.

\section{Weak energy condition}\label{5}
  The weak energy condition states that for all physically reasonable classical matter, as measured by any observer in space-time, must be nonnegative \cite{wald2010general}, i.e.,
  \begin{equation*}
  T_{\mu\nu}\xi^{\mu}\xi^{\nu}\ge0,
  \end{equation*}
  for all timelike $\xi^{\mu}$. With the decomposition of the energy-momentum tensor $T_{\mu\nu}$, the weak energy condition is equivalent to
  \begin{equation}\label{eq:weak energy condition}
  \epsilon\ge0,\quad\epsilon+p_{i}\ge0.
  \end{equation}
  Substituting Eqs. (\ref{eq:line element terms}) and (\ref{eq:state equations}) into Eq. (\ref{eq:weak energy condition}) leads to
  \begin{gather}
  \epsilon=\frac{1}{4\pi}\frac{r^{2}}{\Sigma^{2}}\left[\frac{3g^{2}Mr^{2}}{\left(r^{2}+g^{2}\right)^{\frac{5}{2}}}-\frac{\alpha}{2r}\right]\ge0,\label{eq:energy}\\
  \epsilon+p_{r}=0,\\
  \epsilon+p_{\theta}=\epsilon+p_{\phi}=\frac{2\left(r^{2}-a^{2}\cos^{2}{\theta}\right)\rho'-r\Sigma\rho''}{8\pi\Sigma^{2}}\ge0.
  \end{gather}
\par
\begin{figure}[htbp]
  	\centering
  	\includegraphics[width=.51\textwidth]{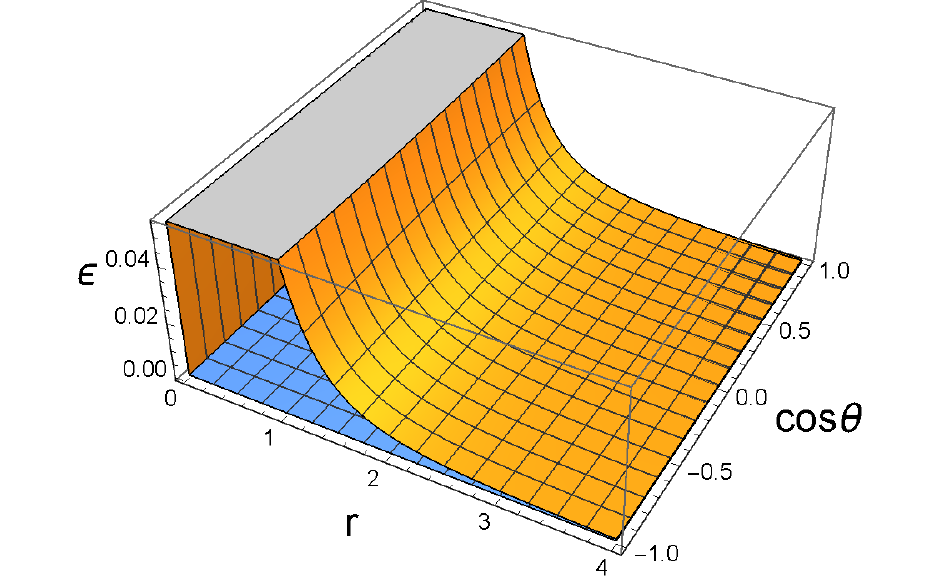}
  	\includegraphics[width=.47\textwidth]{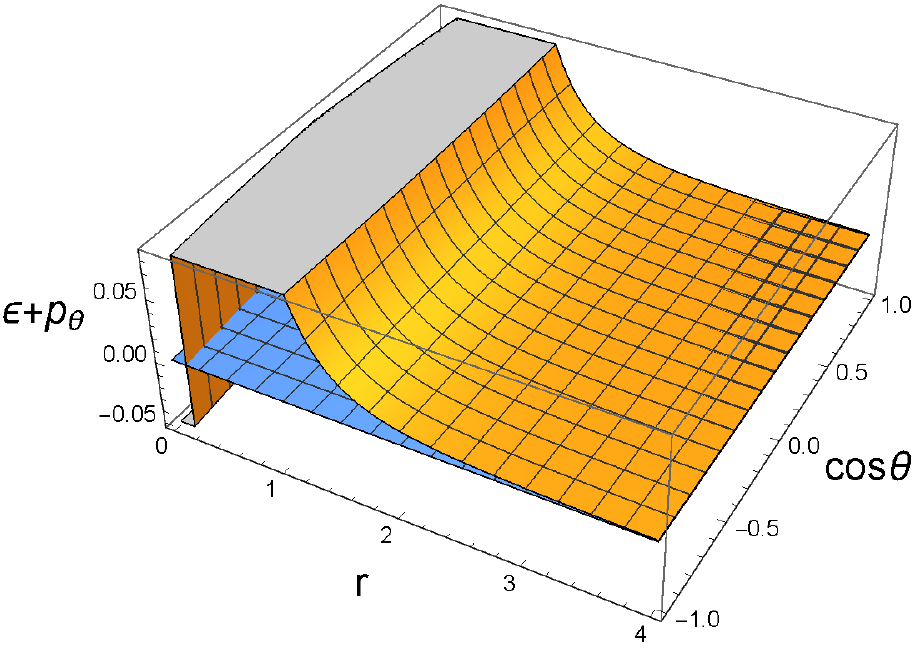}
  	\caption{Dependence of matter density $\epsilon$ and $\epsilon+p_{\theta}$ on radius and angle for a rotating Bardeen BH in PFDM with $M=1, a=0.2, g=0.5, \alpha=-1$.}
  	\label{energy1}
  \end{figure}
\begin{figure}[htbp]
  	\centering
  	\includegraphics[width=.5\textwidth]{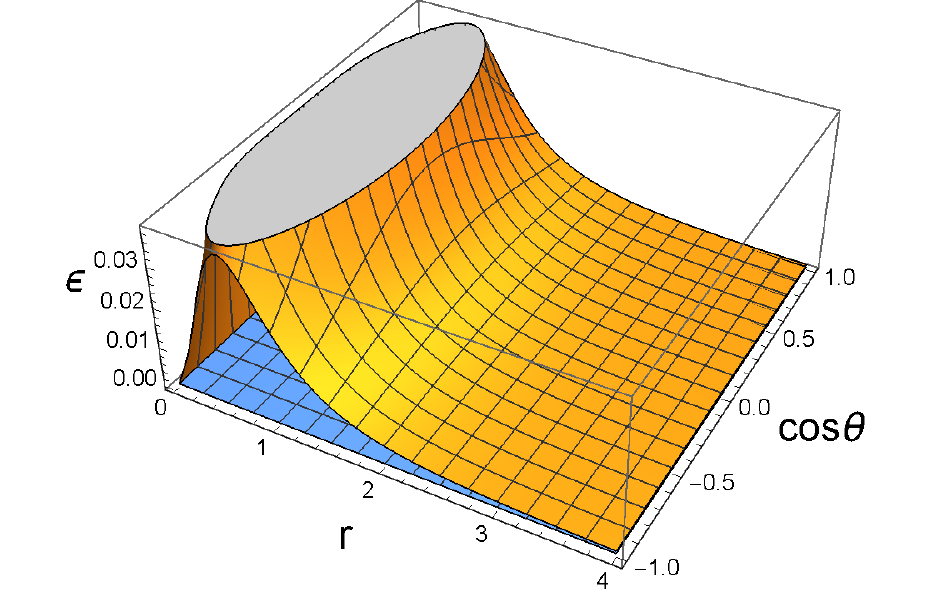}
  	\includegraphics[width=.48\textwidth]{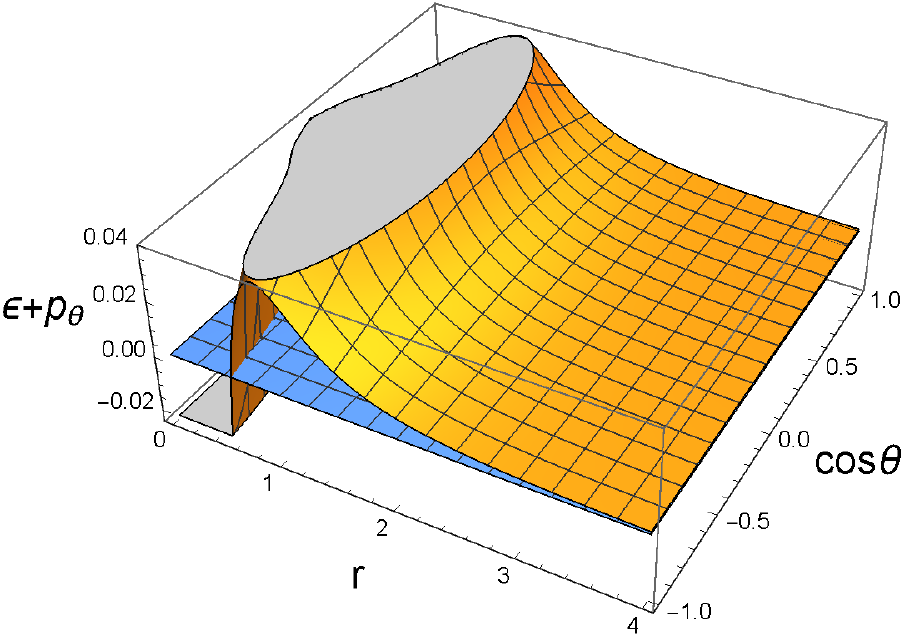}
  	\caption{Dependence of matter density $\epsilon$ and $\epsilon+p_{\theta}$ on radius and angle for a rotating Bardeen BH in PFDM with $M=1, a=0.9, g=0.5, \alpha=-1$.}
  	\label{energy2}
  \end{figure}
\begin{figure}[htbp]
  	\centering
  	\includegraphics[width=.49\textwidth]{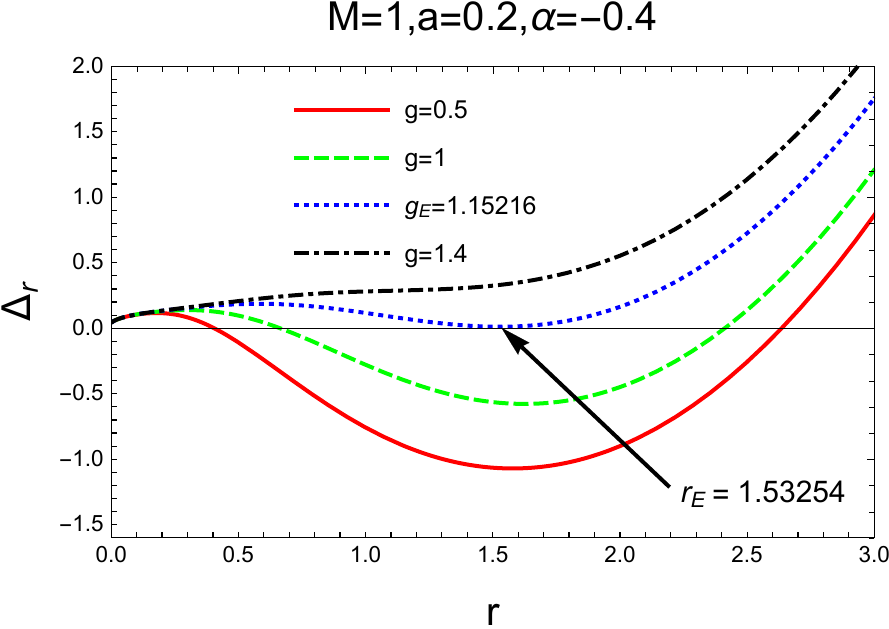}
  	\includegraphics[width=.49\textwidth]{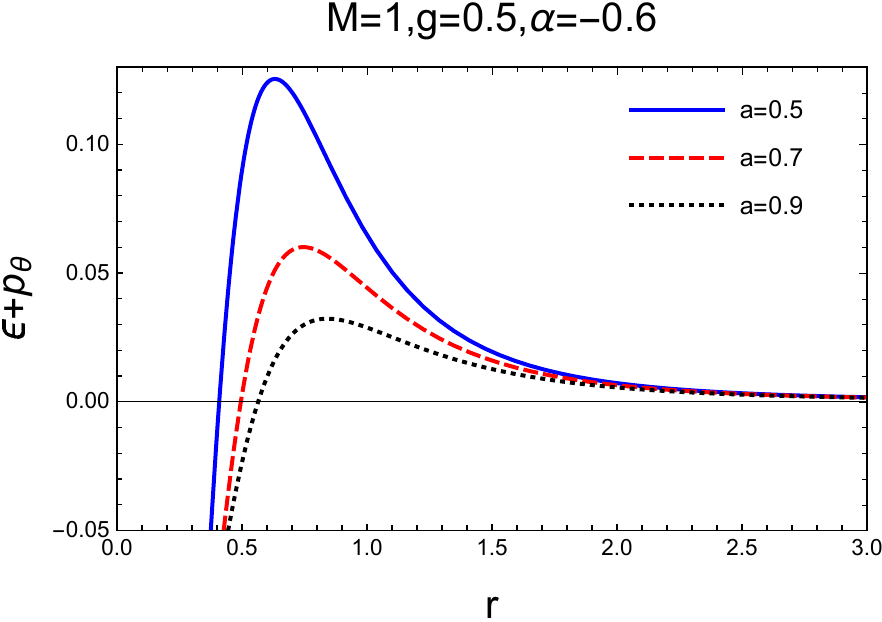}
  	\includegraphics[width=.49\textwidth]{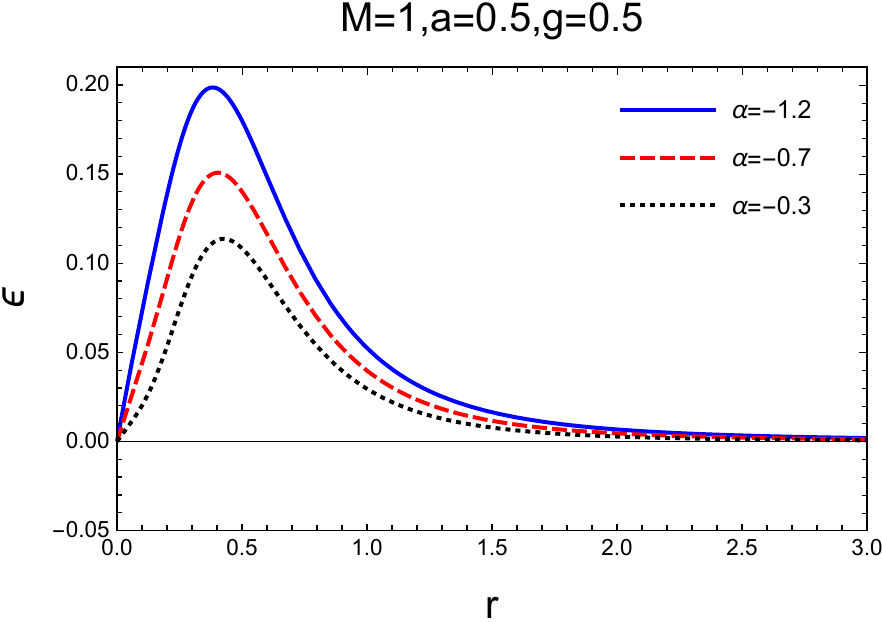}
  	\includegraphics[width=.49\textwidth]{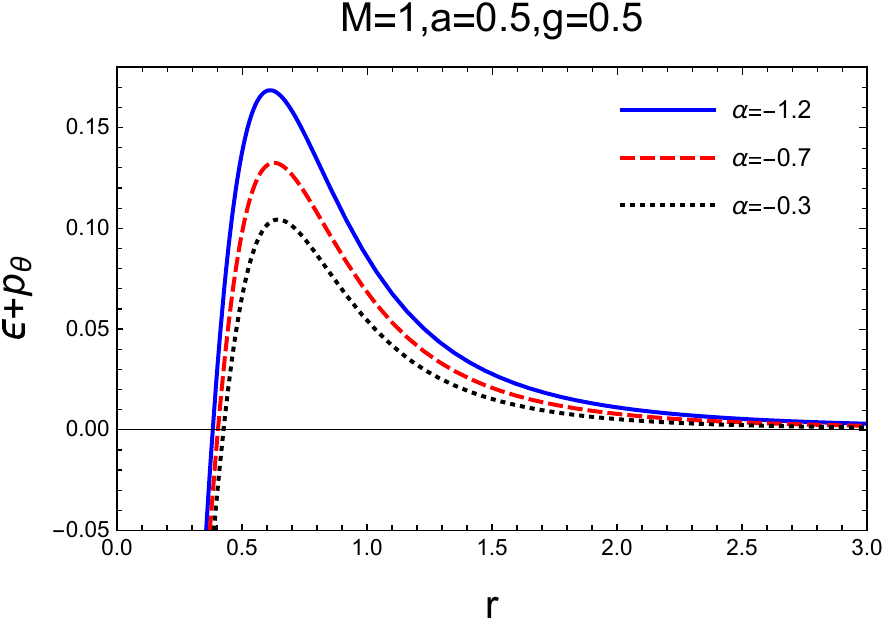}
  	\caption{Plot showing $\epsilon$ and $\epsilon+p_{\theta}$ vs. $r$ for various black hole parameters.}
  	\label{energy3}
  \end{figure}

  According to Eq.~(\ref{eq:state equations}), as examples, we depict the variations of $\epsilon$ and $\epsilon+p_{\theta}/\epsilon+p_{\phi}$ with $r$ and $\cos{\theta}$ under two sets of parameters in Figs. \ref{energy1} and \ref{energy2}. It turns out that the weak energy condition is violated near the origin of the rotating Bardeen in the PFDM, which happens for all the rotating regular black holes \cite{bambi2013rotating,neves2014regular,kumar2019shadow,ghosh2020ergosphere}. In fact, one can realize this by considering the asymptotic behavior of the matter density $\epsilon$ and $\epsilon+p_{\theta}$ near the origin:
  \begin{equation}\label{energybeh1}
  \epsilon\simeq\frac{6Mr^{4}-\alpha g^{3}r}{8\pi g^{3}\left(r^{2}+a^{2}\right)^{2}}\simeq-\frac{\alpha r}{8\pi \left(r^{2}+a^{2}\right)^{2}},\quad r\rightarrow 0,
  \end{equation}
  \begin{equation}\label{energybeh2}
  \epsilon+p_{\theta}\simeq-\frac{3a^{2}Mr^{2}}{2\pi g^{3}\left(r^{2}+a^{2}\right)^{2}}+\frac{\alpha a^{2}}{16\pi r\left(r^{2}+a^{2}\right)^{2}}\simeq\frac{\alpha}{16\pi a^{2}r},\quad r \rightarrow 0,
  \end{equation}
  where we set $\cos{\theta}=\pm 1$. Eq. (\ref{energybeh2}) implies that, for the rotating Bardeen black hole ($\alpha=0$), the violation of weak energy cannot be prevented if $a\neq 0$. As a verification of Eqs. (\ref{energybeh1}) and (\ref{energybeh2}), the dependence of  $\epsilon$ and $\epsilon+p_{\theta}$ on $a$ and $\alpha$ is plotted in Fig. \ref{energy3}. Further more, at large $r$, the matter density $\epsilon$ and $\epsilon+p_{\theta}$ behave as
  \begin{equation}\label{weakinf1}
  \epsilon\simeq\frac{3g^{2}M}{4\pi r\left(r^{2}+a^{2}\right)^{2}}-\frac{\alpha r}{8\pi\left(r^{2}+a^{2}\right)^{2}}\simeq-\frac{\alpha}{8\pi r^{3}},\quad r\rightarrow \infty,
  \end{equation}
  \begin{equation}\label{weakinf2}
  \epsilon+p_{\theta}\simeq\frac{15Mg^{2}}{8\pi r^{5}}-\frac{3\alpha}{16\pi r^{3}}\simeq-\frac{3\alpha}{16\pi r^{3}},\quad r\rightarrow\infty
  \end{equation}
  for $\cos{\theta}=\pm 1$. From Eqs. (\ref{weakinf1}) and (\ref{weakinf2}),  one can find that for the Bardeen black hole in PFDM, if $\alpha>0$, then both $\epsilon$ and $\epsilon+p_{\theta}$ are negative at the large distances, which is quite unreasonable from the perspective of observation. Therefore, we always assume $\alpha<0$ in the following discussions.


\section{Properties of rotating Bardeen black hole in perfect fluid dark matter}\label{6}
\subsection{Horizons}
  Similar to the Kerr black hole, the space-time metric (\ref{eq:rotating metric}) is singular at $\Delta_{r}=0$, which corresponds to the horizons of the rotating black hole. In other words, the horizons of the rotating Bardeen black hole in PFDM are solutions of
  \begin{equation}\label{eq:horziom}
  \Delta_{r}=r^{2}+a^{2}-\frac{2Mr^{4}}{\left(r^{2}+g^{2}\right)^{\frac{3}{2}}}+\alpha r\ln{\frac{r}{\vert\alpha\vert}}=0.
  \end{equation}
\par
\begin{figure}[htbp]
  	\centering
  	\subfigure{
  		\includegraphics[width=.49\textwidth]{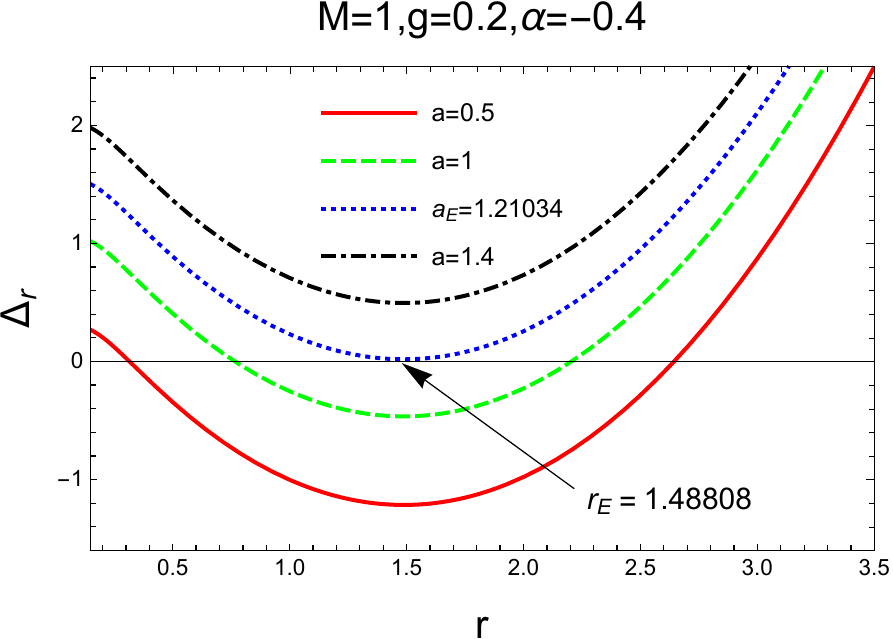}
  		\includegraphics[width=.49\textwidth]{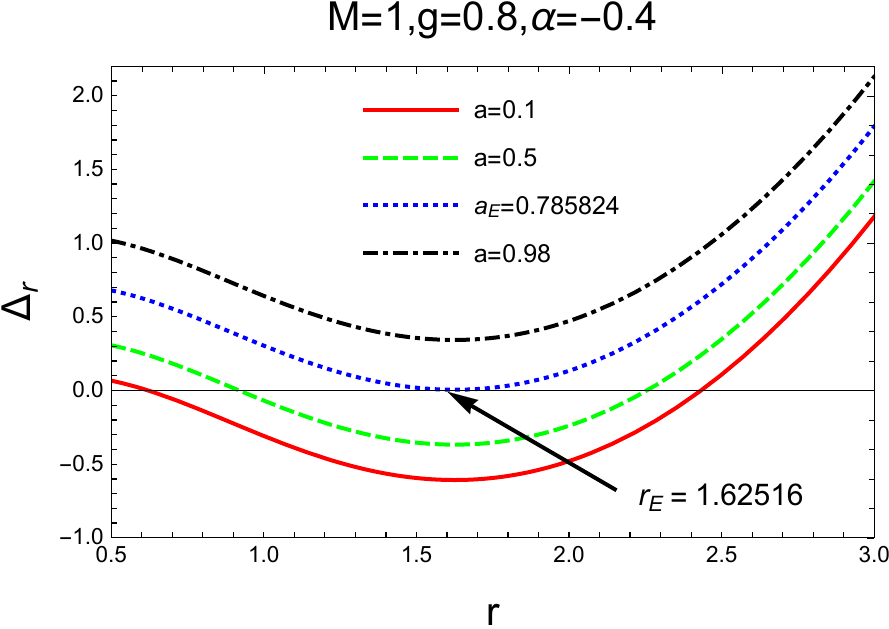}
  	}
  	\subfigure{
  		\includegraphics[width=.49\textwidth]{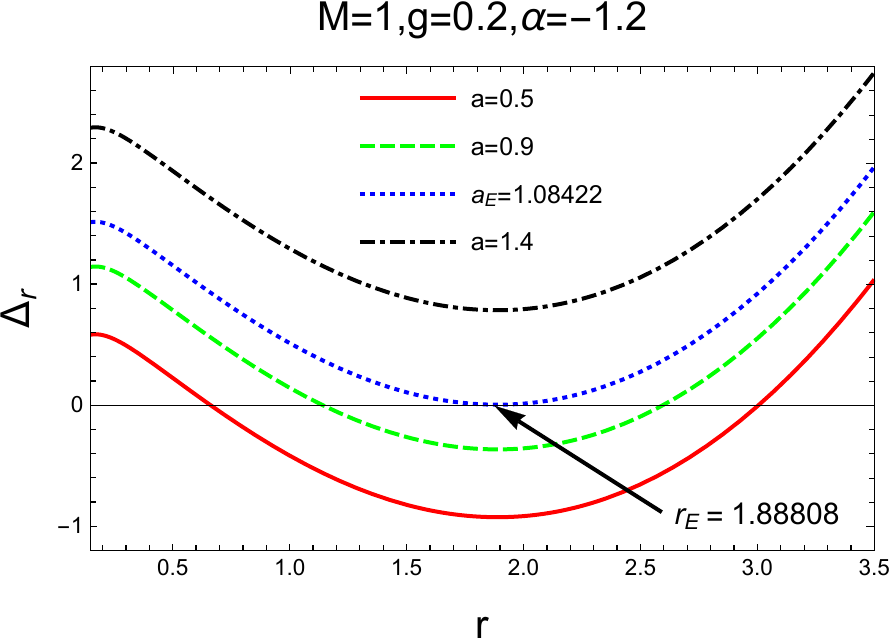}
  		\includegraphics[width=.49\textwidth]{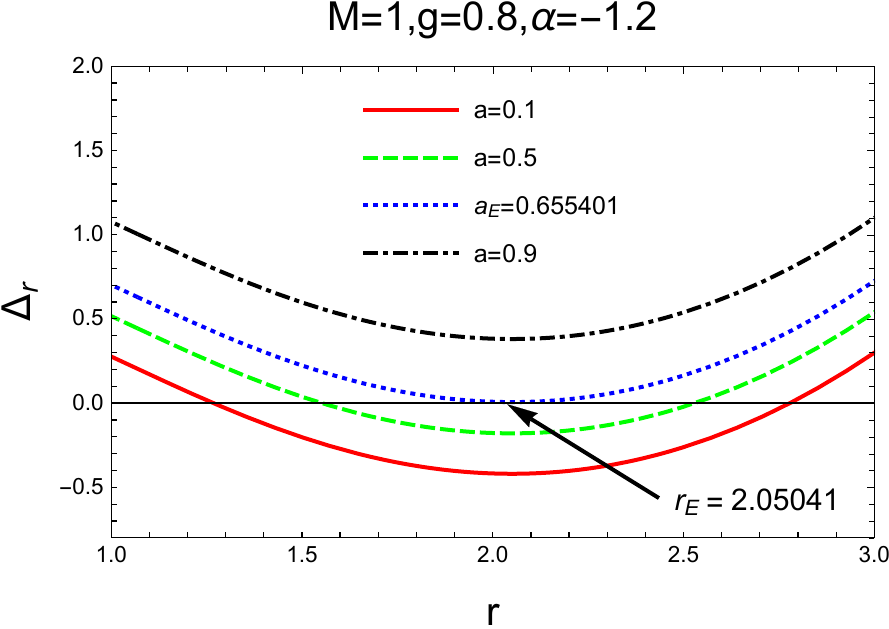}
  	}
  	\caption{Plot showing the behavior of horizons vs. $r$ for a set of fixed values of $M=1, g$, and $\alpha$ by varying $a$.}
  	\label{fig:horizon1}
  \end{figure}
\begin{figure}[htbp]
  	\centering
  	\subfigure{
  		\includegraphics[width=.49\textwidth]{BHag1}
  		\includegraphics[width=.49\textwidth]{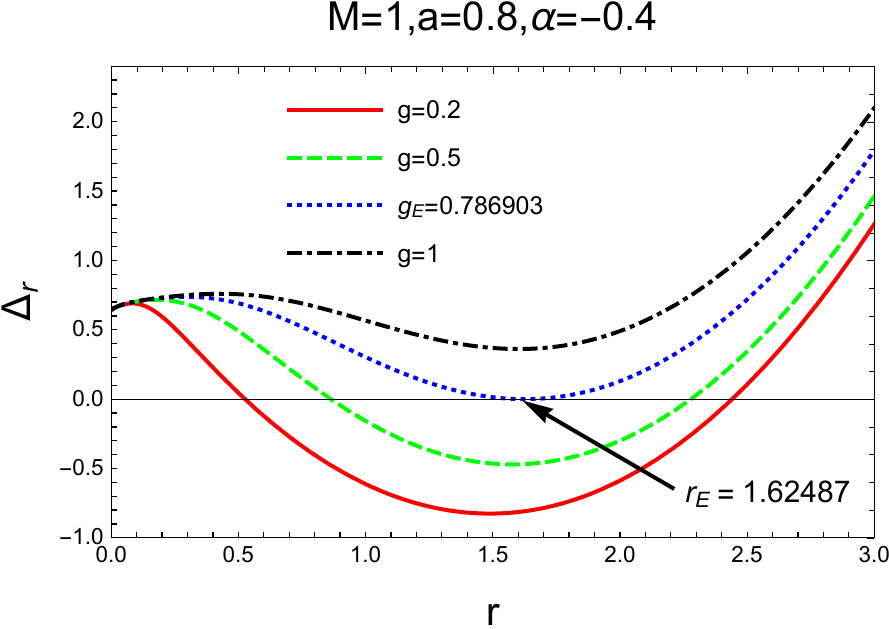}
  	}
  	\subfigure{
  		\includegraphics[width=.49\textwidth]{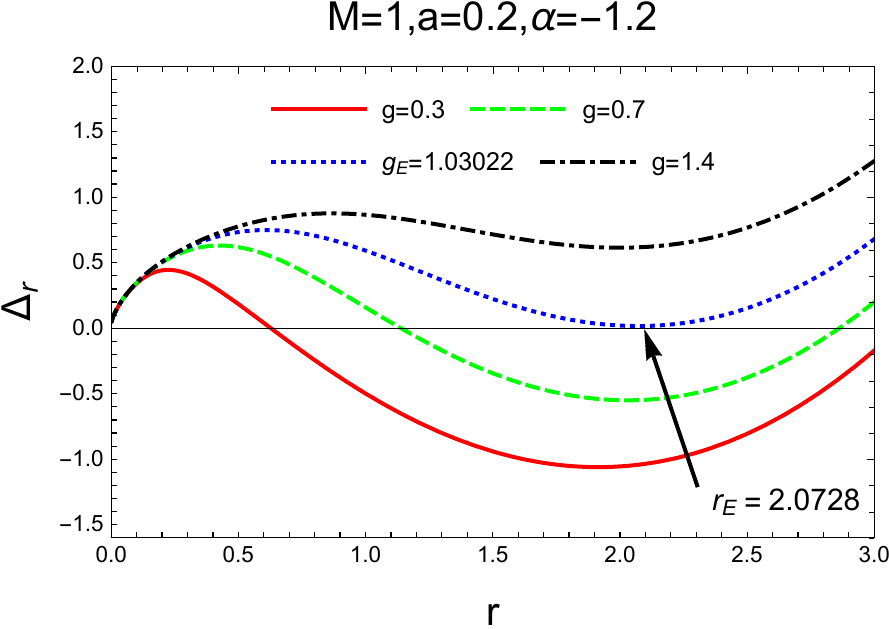}
  		\includegraphics[width=.49\textwidth]{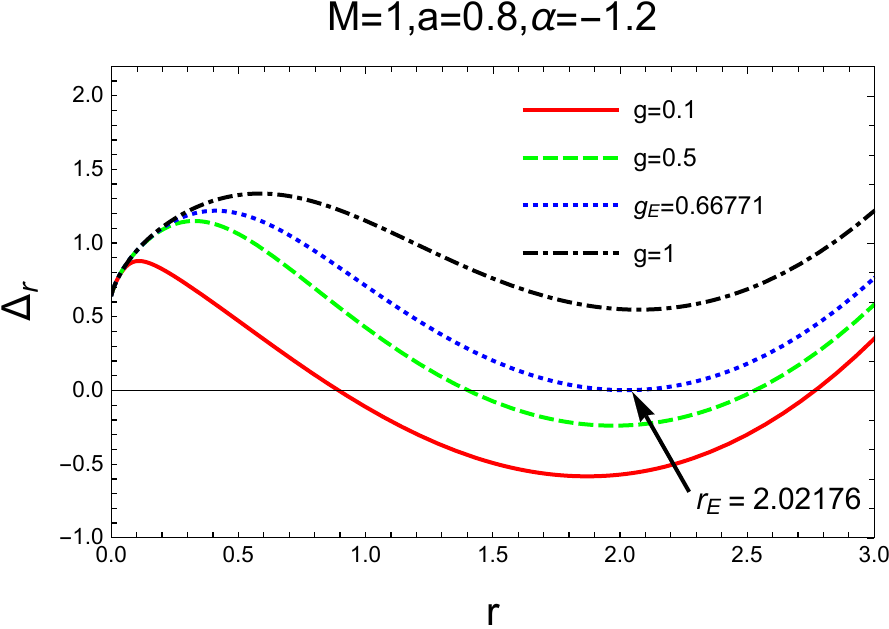}
  	}
  	\caption{Plot showing the behavior of horizons vs. $r$ for a set of fixed values of $M=1, a$, and $\alpha$ by varying $g$.}
  	\label{fig:horizon2}
  \end{figure}

  Obviously, the radii of horizons depend on the rotation parameter $a$, magnetic charge $g$ and dark matter $\alpha$. The numerical analysis of Eq.~(\ref{eq:horziom}) suggests the possibility of two roots for a set of values of parameters, which correspond the Cauchy horizon $r_{-}$ (smaller root) and the event horizon $r_{+}$ (larger root), respectively. The variation of $\Delta_{r}$ with respect to $r$ for the different values of parameters $a$, $g$, and $\alpha$ is depicted in Figs.~\ref{fig:horizon1} and \ref{fig:horizon2}. As can be seen from Fig.~\ref{fig:horizon1}, for any fixed parameters $g$ and $\alpha$, when $a<a_{E}$, the radii of Cauchy horizons increase with the increasing $a$ while the radii of event horizons decrease with $a$. For $a=a_{E}$, these two horizons meet at $r_{E}$, that is, we have an extremal black hole with degenerate horizons. Here, the critical rotation parameter $a_{E}$ and the corresponding critical radius $r_{E}$ can be obtained by combining $\Delta_{r}=0$ with $\partial_r\Delta_{r}=0$. If $a>a_{E}$, Eq.~(\ref{eq:horziom}) has no root, i.e., no horizon exists, which means there is no black hole. A similar analysis can be applied to Fig.~\ref{fig:horizon2}. The result shows that, for any given values of parameters $a$ and $\alpha$, two horizons get closer first with the increase of $g$, then coincide when $g=g_{E}$ and eventually disappear.

\par
\begin{figure}[htbp]
  	\centering
  	\subfigure[~Rotating Bardeen black hole in PFDM]{
  		\includegraphics[width=.47\textwidth]{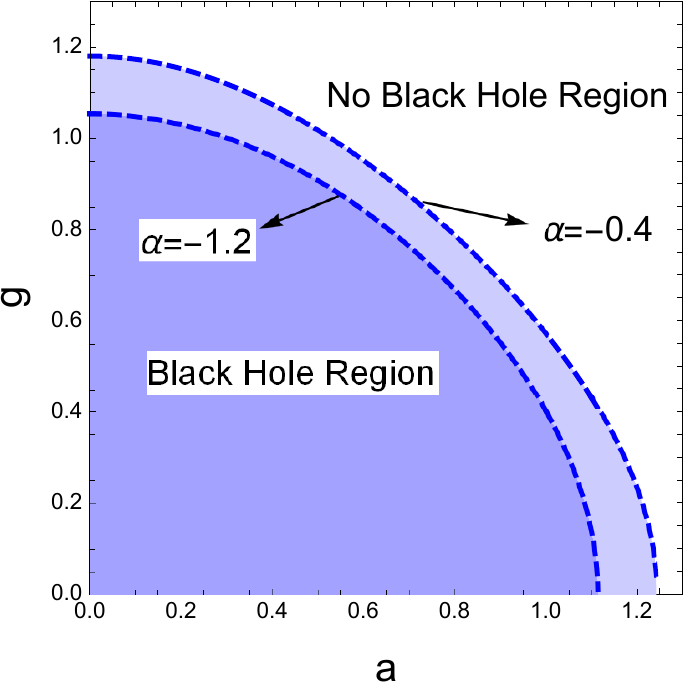}
  		\label{fig:region1}
  	}
  	\subfigure[~Kerr-Newman black hole]{
  		\includegraphics[width=.47\textwidth]{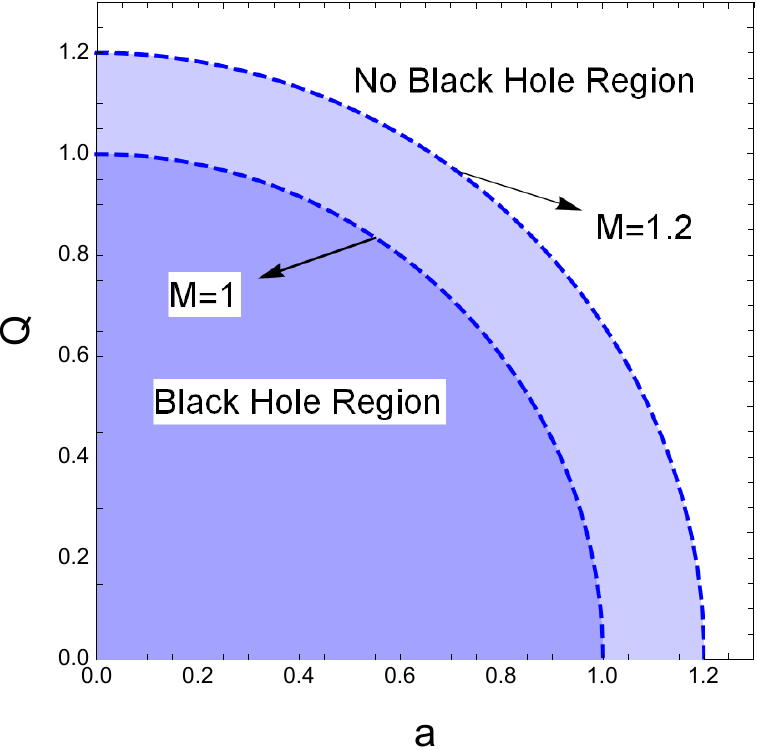}
  		\label{fig:region2}
  	}
  	\caption{(a)~The parameter space ($a, g$) for various values of $\alpha=-0.4,-1.2$. (b)~The parameter space ($a, Q$) for various values of $M=1,1.2$.}
  	\label{blackholeregion1}
  \end{figure}

  Next, we further analyze the behavior of the horizons of rotating Bardeen black holes in PFDM. As shown in Fig.~\ref{fig:horizon1}, when the dark matter parameter $\alpha$ is fixed, for any given magnetic charge $g$, there always exists a critical value of $a_{E}$ at which the two horizon coincide. Varying $g$, we thus obtained a critical curve with $M=1$ in the parameter space ($a, g$) (see Fig.~\ref{fig:region1}), and every point on the curve corresponds to an extremal black hole with the degenerate horizons. It is worth noting that the critical curve separates the black hole region from the no black hole region. Fig.~\ref{fig:region1} implies that, the extremal value of the rotation parameter $a_{E}$ decreases with increasing magnetic charge $g$. As a comparison, the critical curve of Kerr-Newman black holes is also depicted in Fig.~\ref{fig:region2}. Actually, one can give the analytic expression of the curve, $a^{2}+Q^{2}=M^{2}$ with $Q$ denoting the total charge of black holes. From this expression as well as Fig.~\ref{fig:region2}, it is easily find that the critical curve is a part of circle.

\subsection{Ergosphere}
\begin{figure}
  	\centering
  	\includegraphics[width=0.24\textwidth]{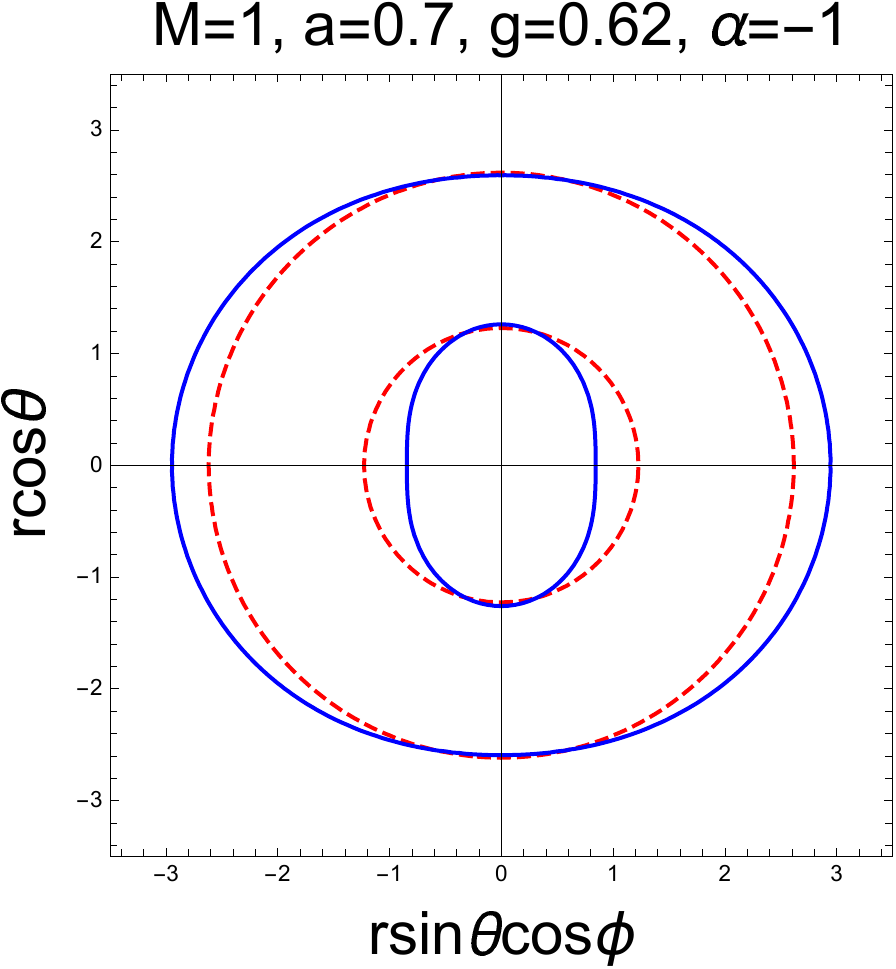}
  	\includegraphics[width=0.24\textwidth]{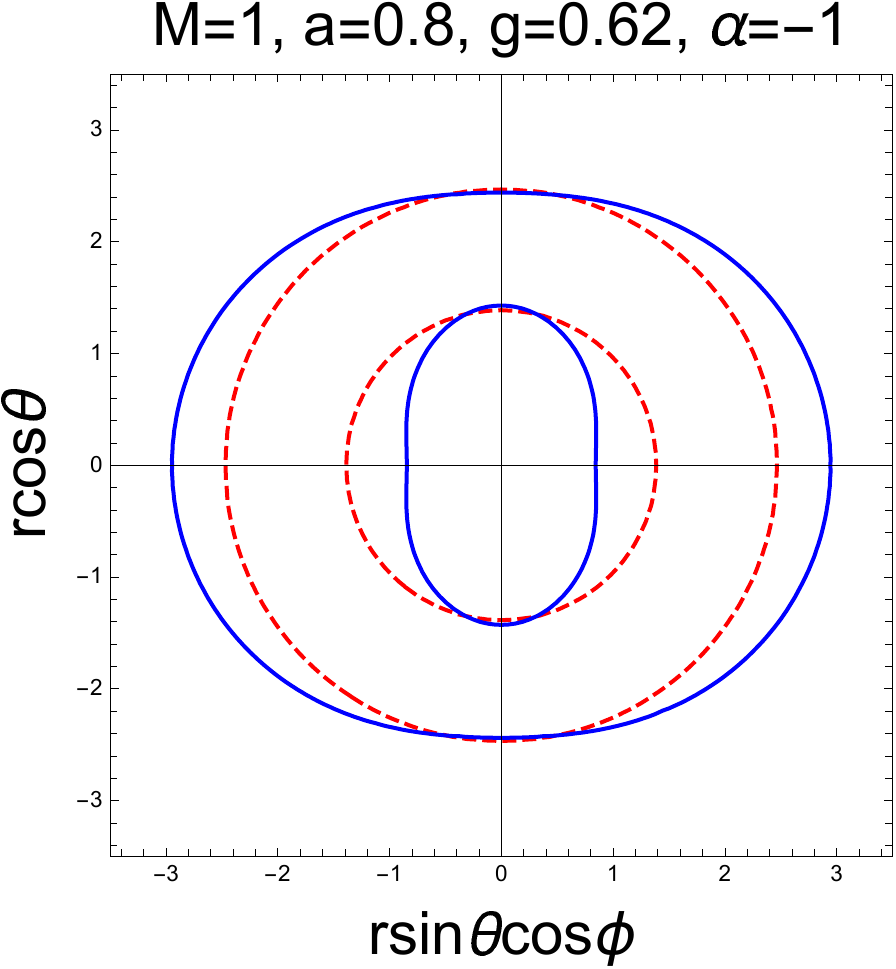}
  	\includegraphics[width=0.24\textwidth]{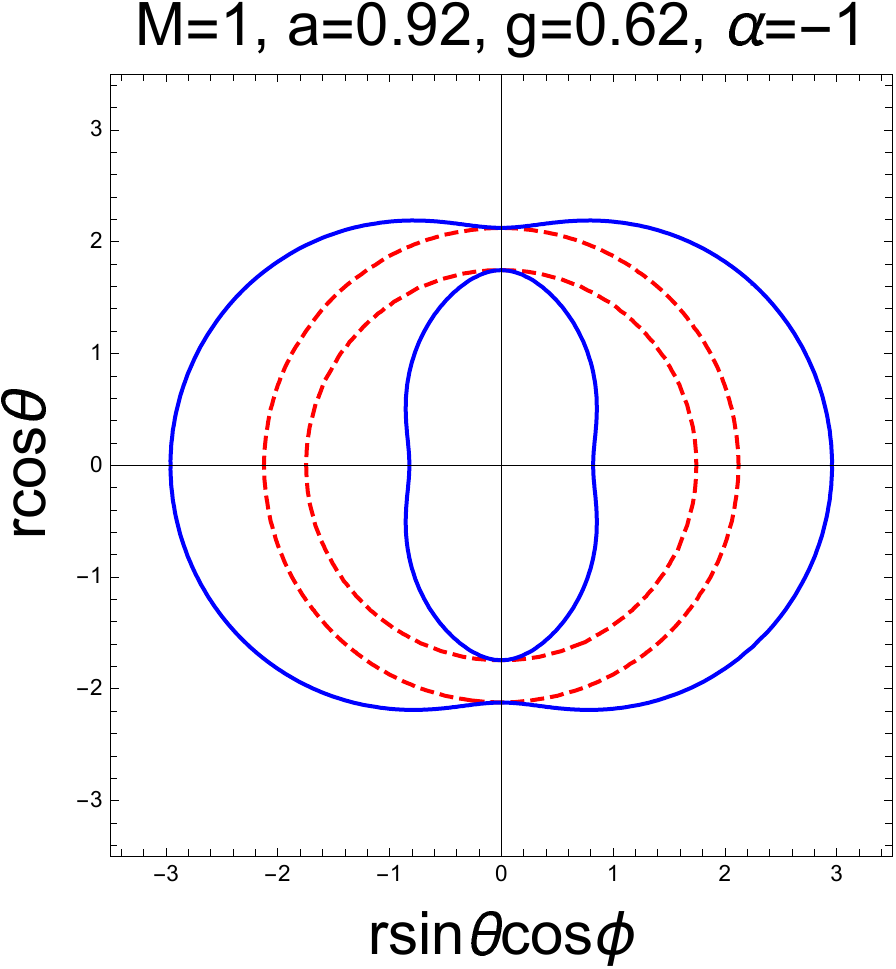}
  	\includegraphics[width=0.24\textwidth]{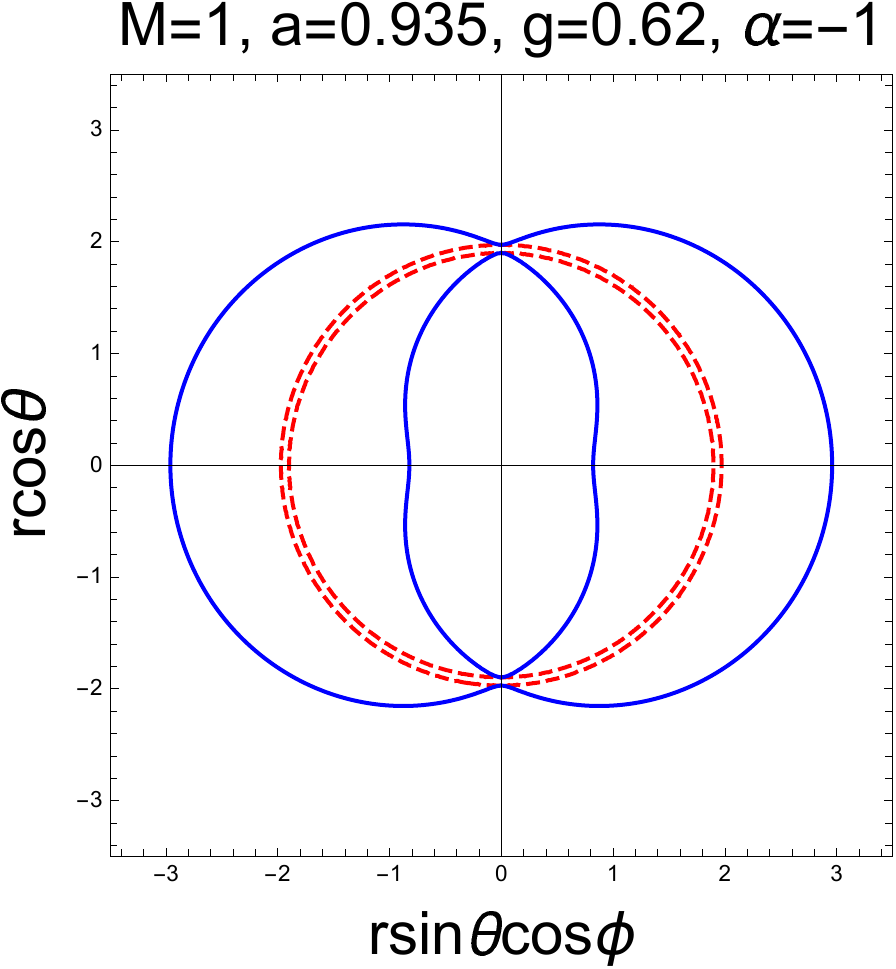}
  	\caption{Illustration of the shapes of ergospheres and horizons with different value of $a, g, \alpha$. The blue solid lines and red dashed lines correspond to stationary limit surfaces and horizons, respectively.}
  	\label{fig:ergosphere}
  \end{figure}

  The ergosphere is a region bounded by the event horizon $r_{+}$ and the outer stationary limit surface (denoted by $r_{+}^{S}$), in fact, it lies outside the black hole. Interestingly, the ergosphere can be used to extract energy from a rotating black hole, which is known as the Penrose process \cite{wald2010general}. We will discuss it in detail in the next section. The stationary limit surface, that is, infinite redshift surface, is a surface where the time-translation Killing vector $K^\mu=\partial_t$ satisfy $K^{\mu}K_{\mu}=0$, or equivalently,
  \begin{equation}\label{ergosphere}
  r^{2}+a^{2}\cos^{2}{\theta}-\frac{2Mr^{4}}{\left(r^{2}+g^{2}\right)^{\frac{3}{2}}}+\alpha r\ln{\frac{r}{\vert\alpha\vert}}=0.
  \end{equation}
  Solving Eq. (\ref{ergosphere}) for various values of the parameters numerically, one can generally get two roots, i.e. inner stationary limit surface $r_{-}^{S}$ and outer stationary limit surface $r_{+}^{S}$. Fig. \ref{fig:ergosphere} shows the shapes of the ergospheres and horizons for the rotating Bardeen black hole surrounded by perfect fluid dark matter. It can be seen that the size of the ergosphere increases with the rotation parameter $a$ (see Fig.~\ref{fig:ergosphere} horizontally) and increases slightly with the increase of $g$ (see Fig.~\ref{fig:ergosphere} from the top down). Moreover, one can find that there exists a critical value $a_{E}$ at which the inner horizon and outer horizon degenerate into one, when $a>a_{E}$, the ergosphere disappears.

  \begin{figure}[htbp]
  	\centering
  		\includegraphics[width=0.24\textwidth]{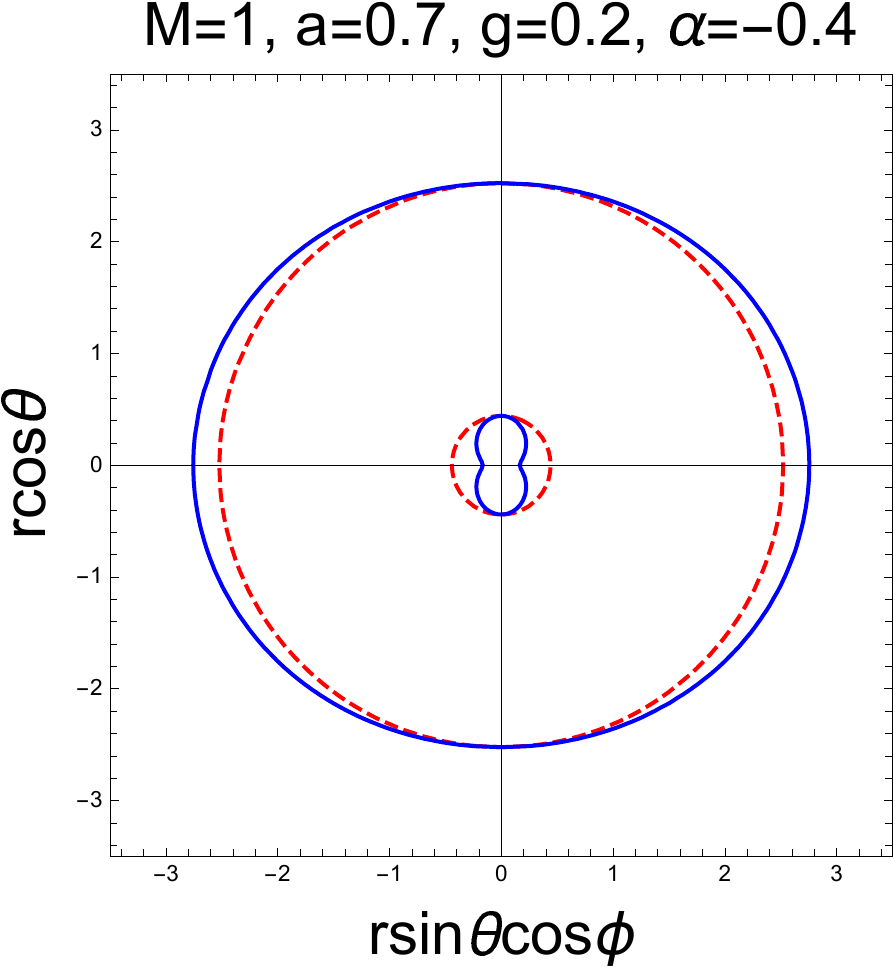}
  		\includegraphics[width=0.24\textwidth]{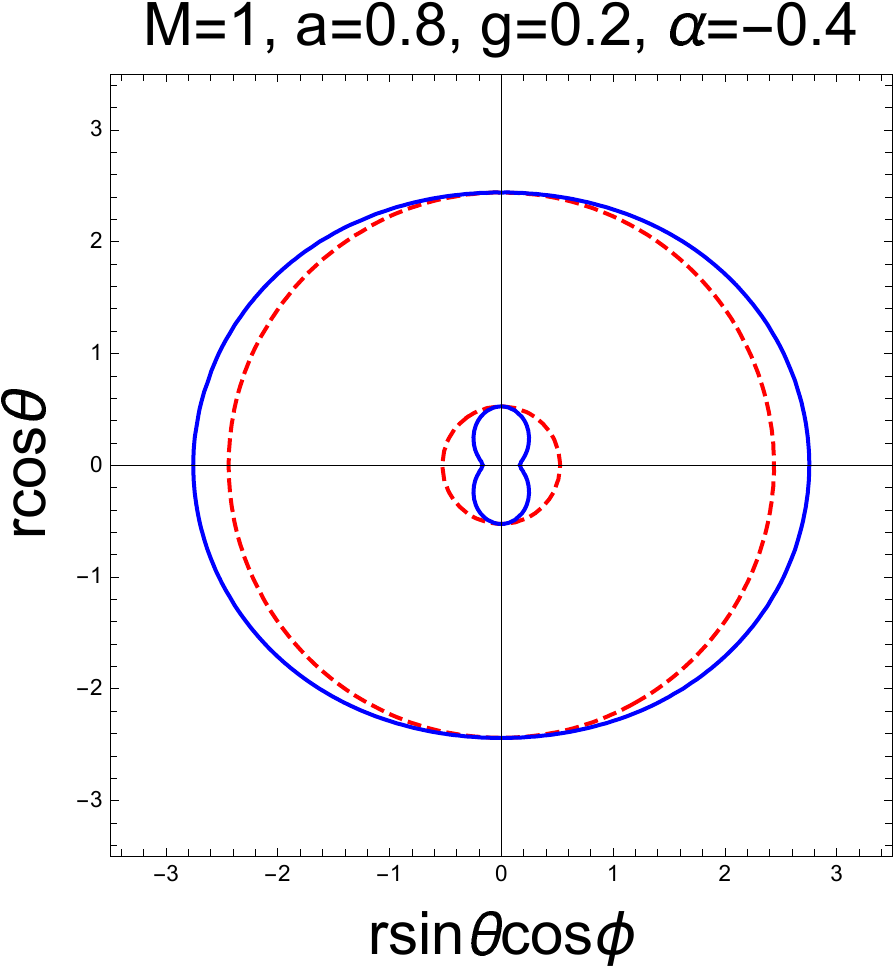}
  		\includegraphics[width=0.24\textwidth]{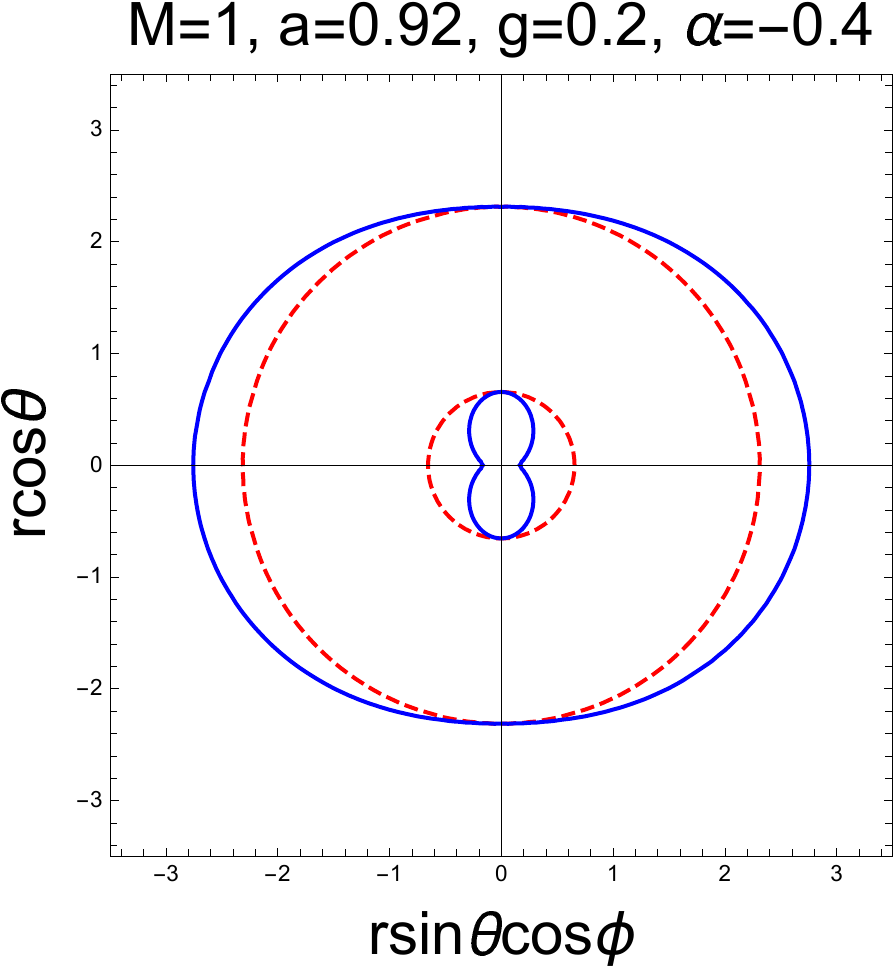}
  		\includegraphics[width=0.24\textwidth]{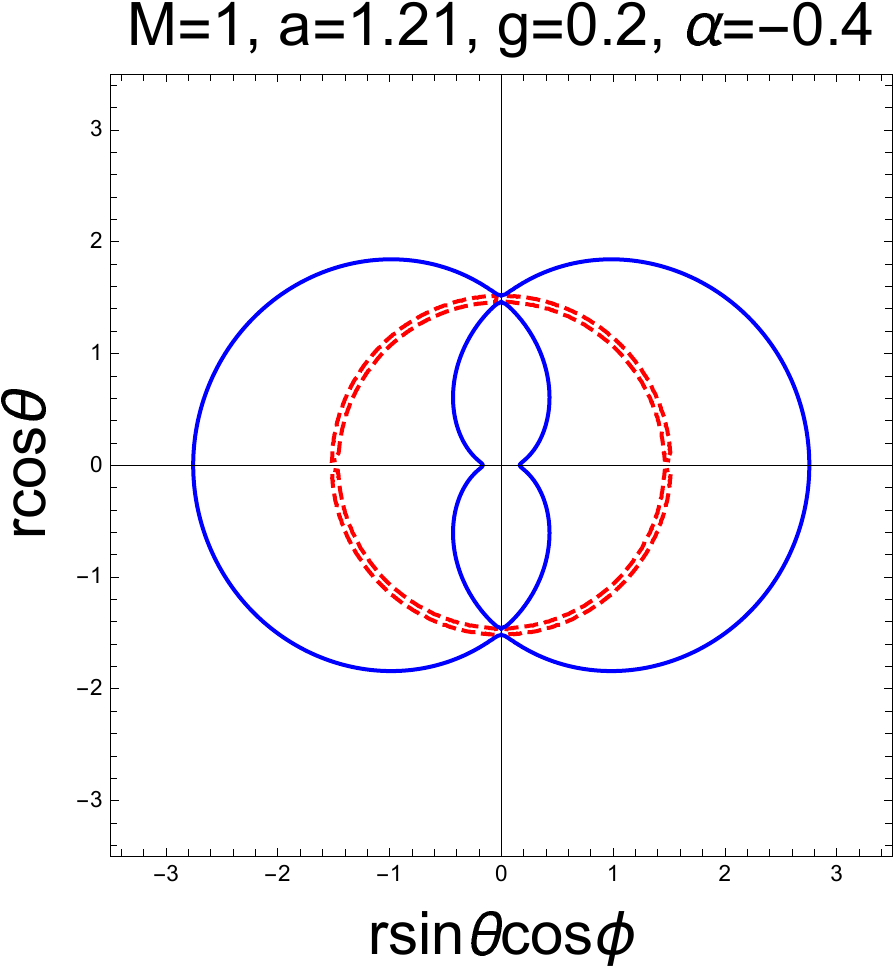}
  		\includegraphics[width=0.24\textwidth]{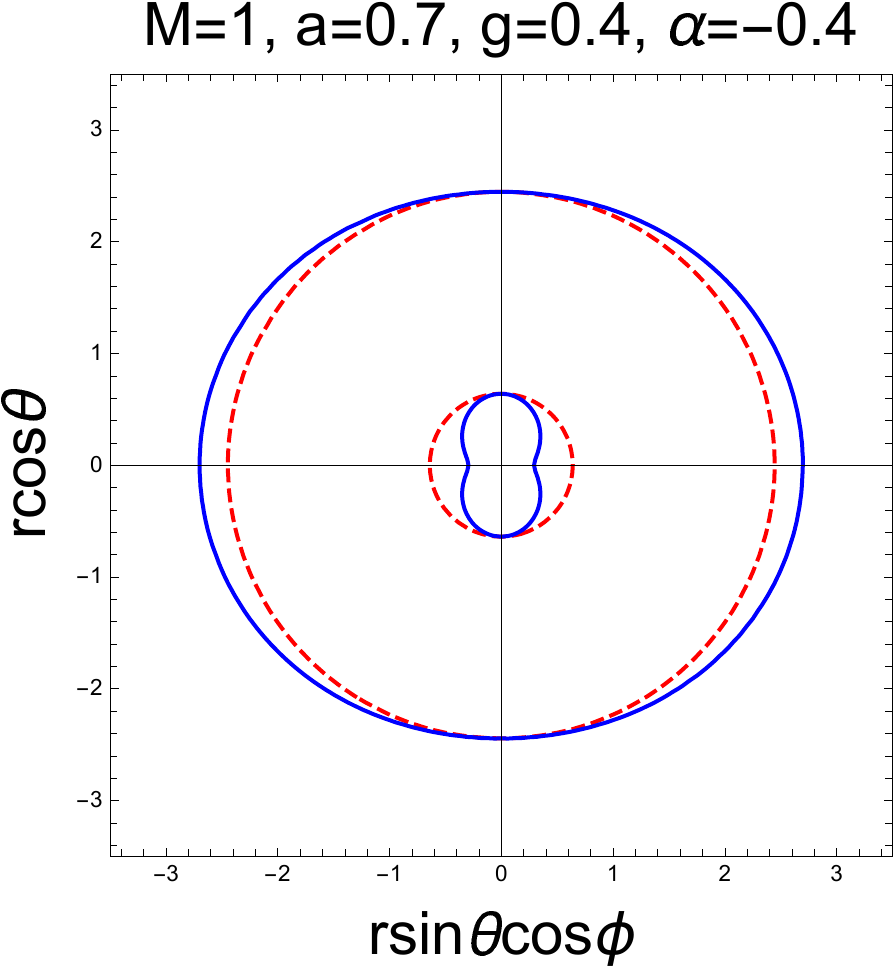}
  		\includegraphics[width=0.24\textwidth]{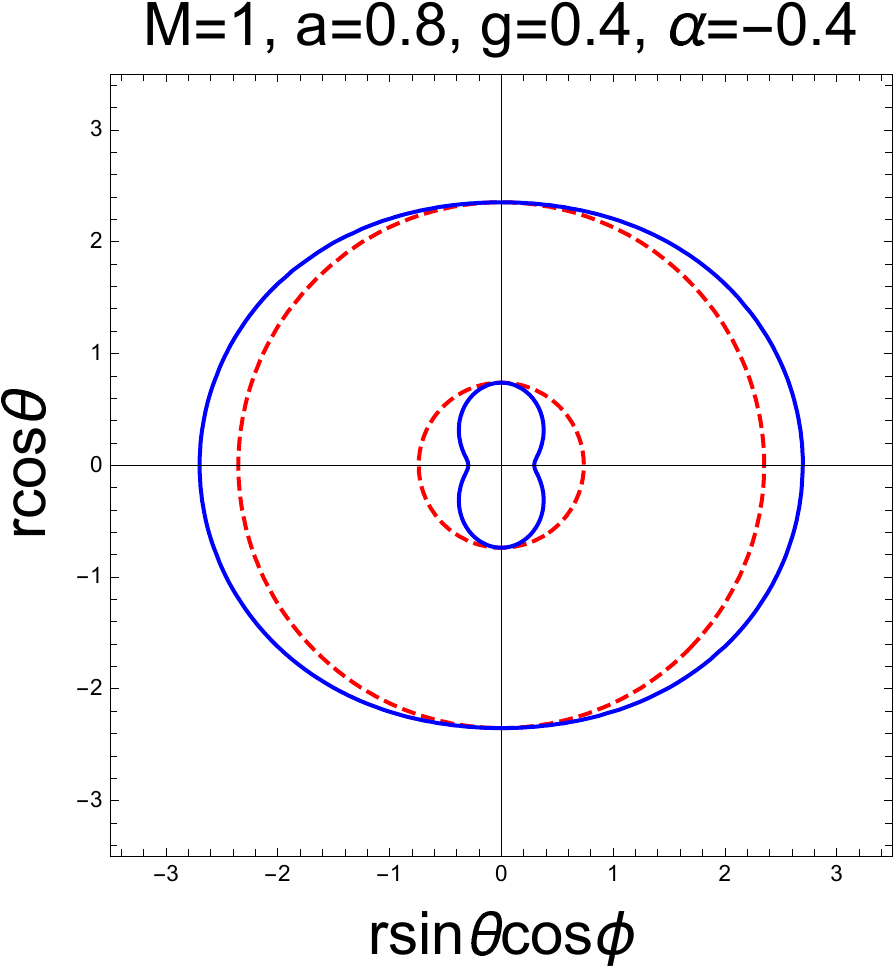}
  		\includegraphics[width=0.24\textwidth]{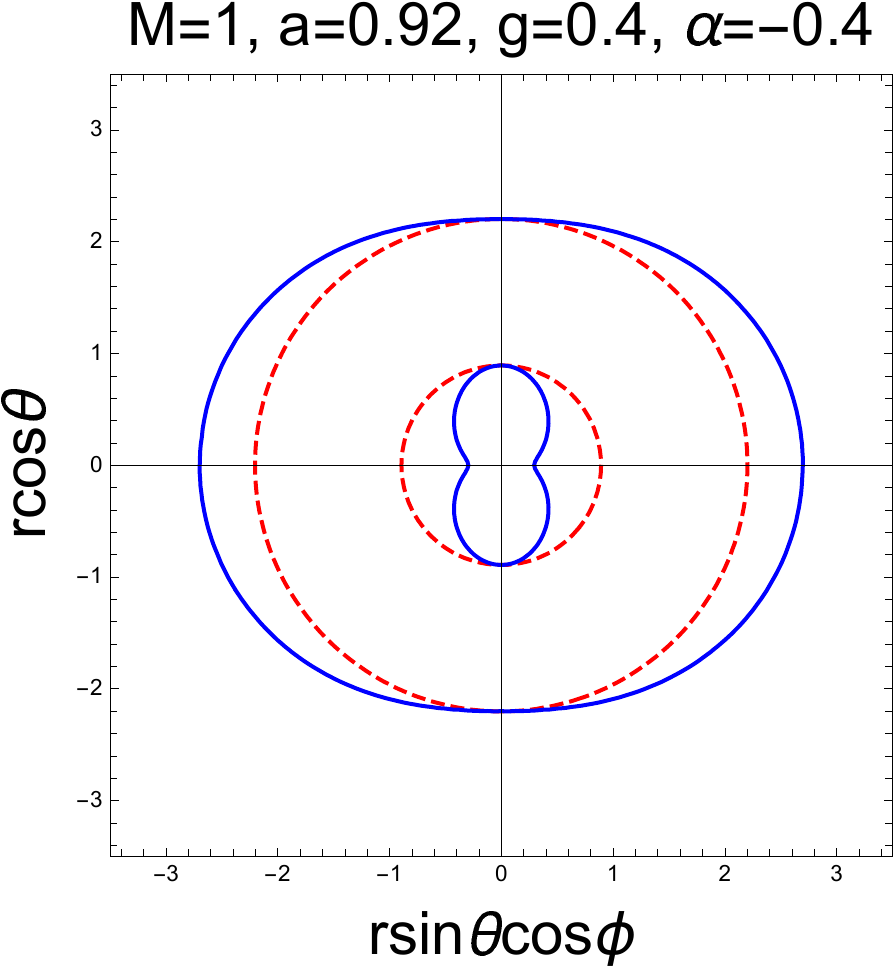}
  		\includegraphics[width=0.24\textwidth]{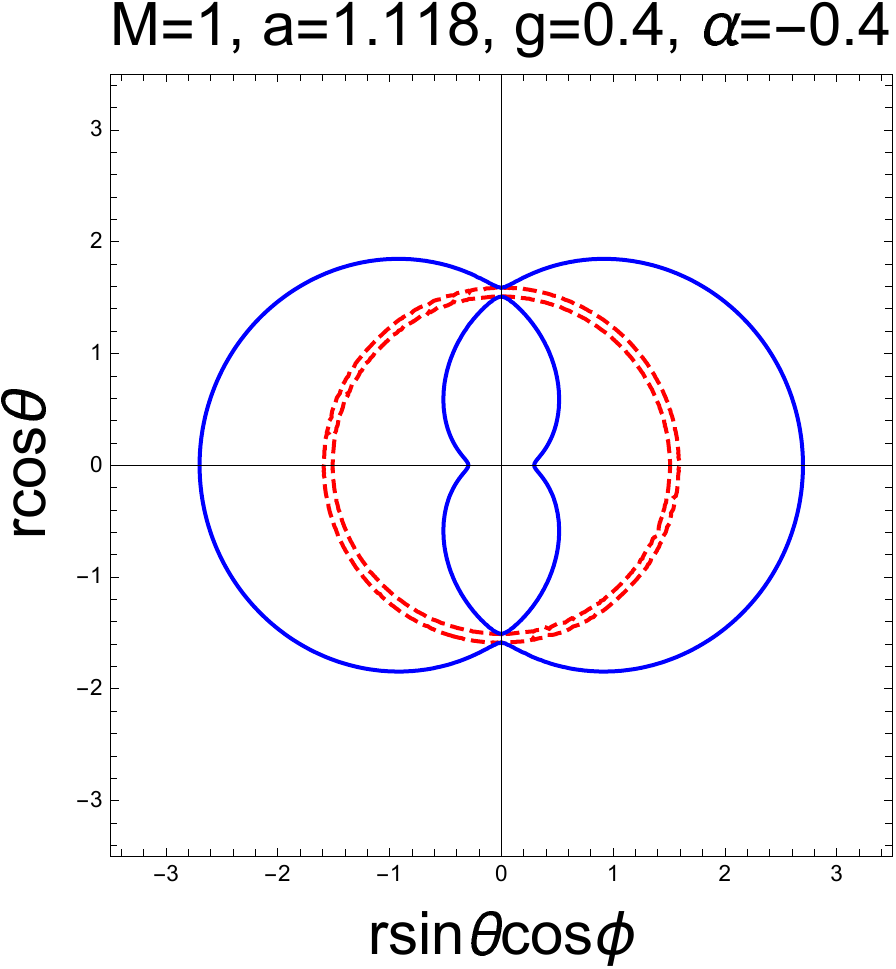}
  		\includegraphics[width=0.24\textwidth]{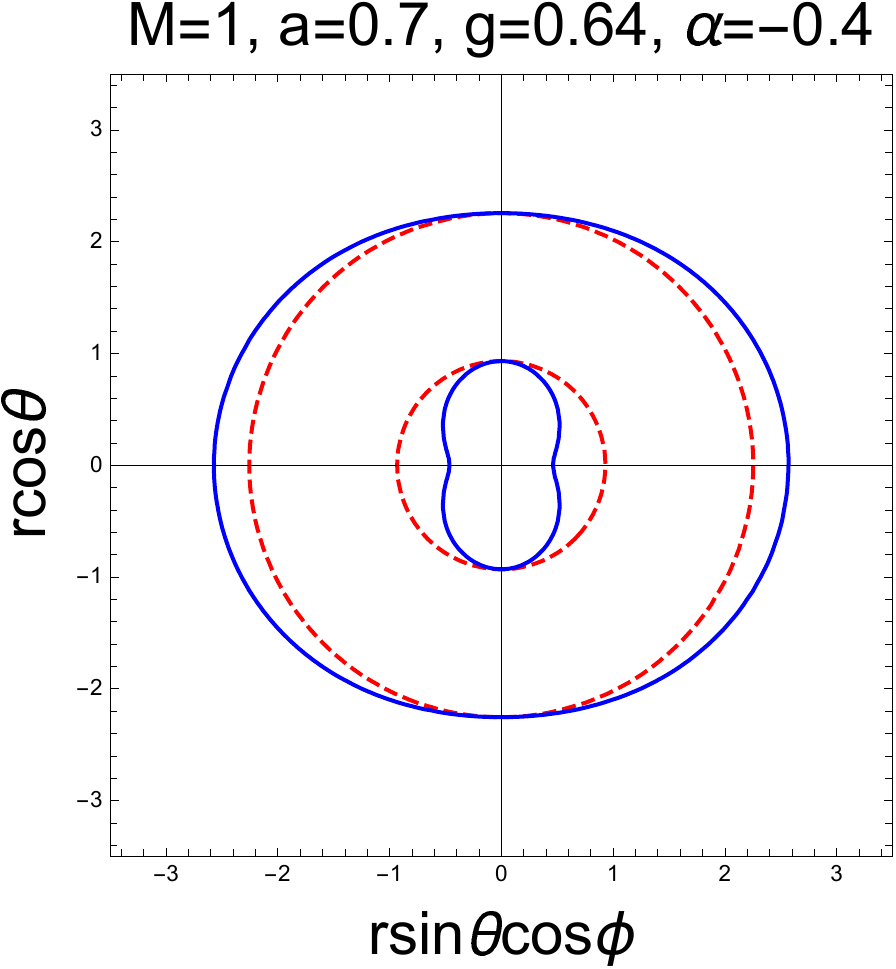}
  		\includegraphics[width=0.24\textwidth]{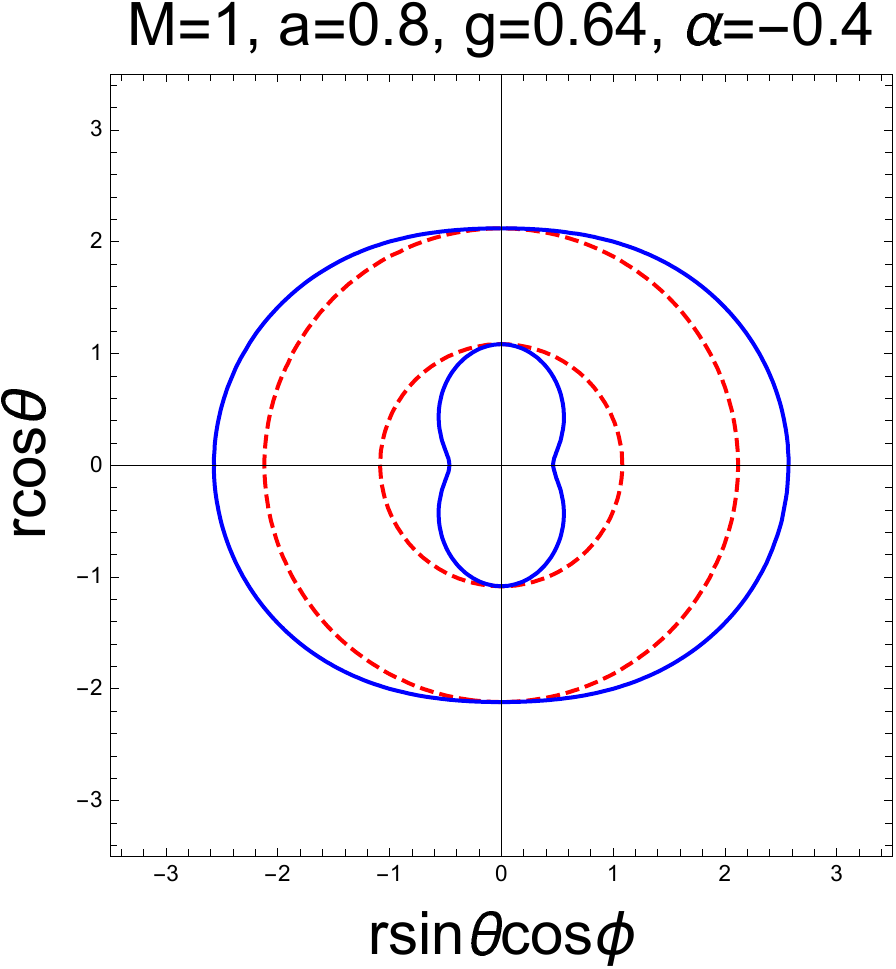}
  		\includegraphics[width=0.24\textwidth]{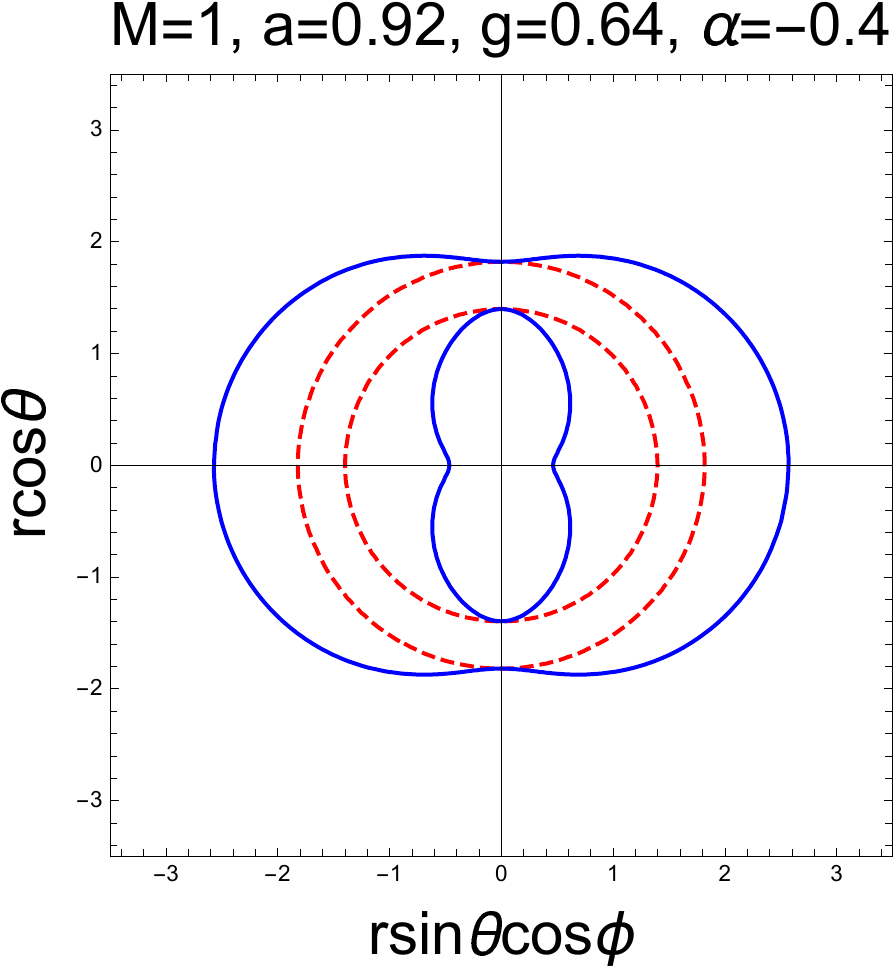}
  		\includegraphics[width=0.24\textwidth]{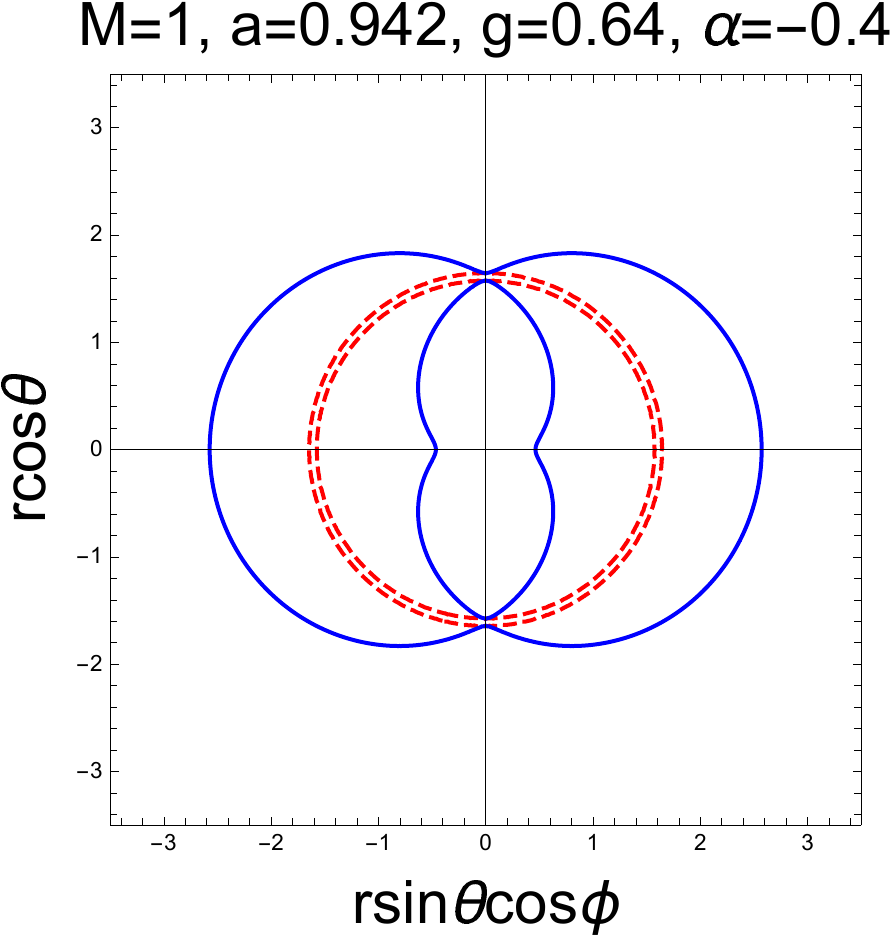}
  		\includegraphics[width=0.24\textwidth]{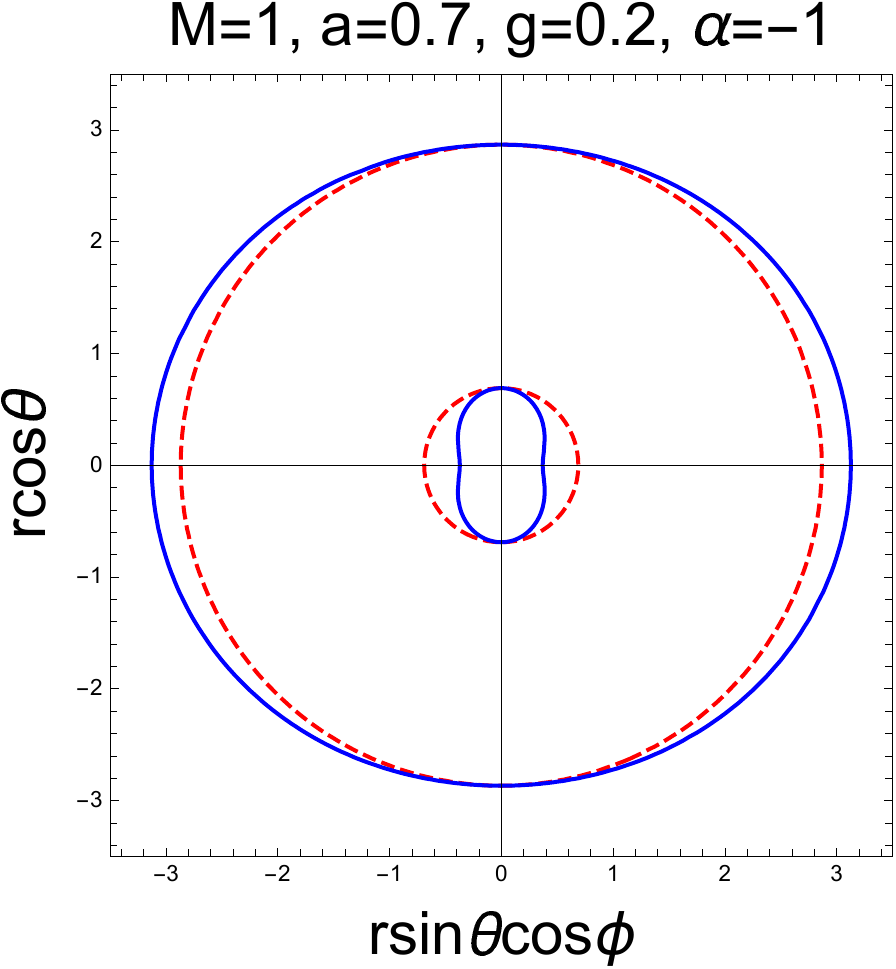}
  		\includegraphics[width=0.24\textwidth]{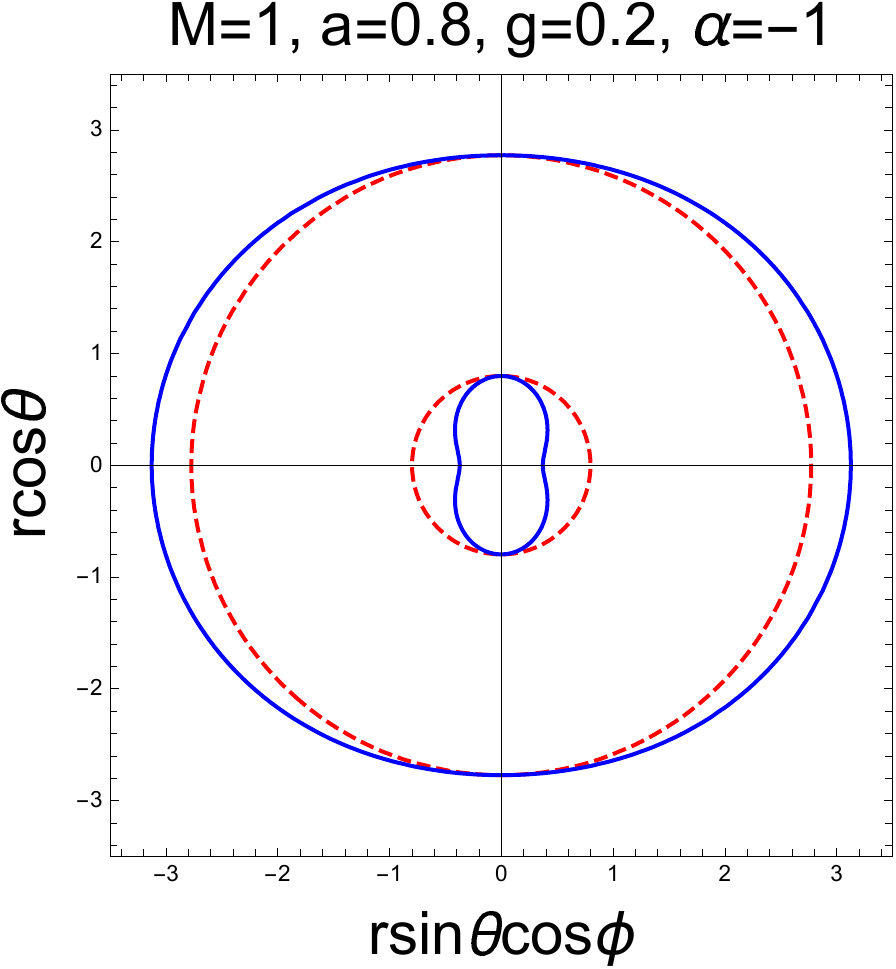}
  		\includegraphics[width=0.24\textwidth]{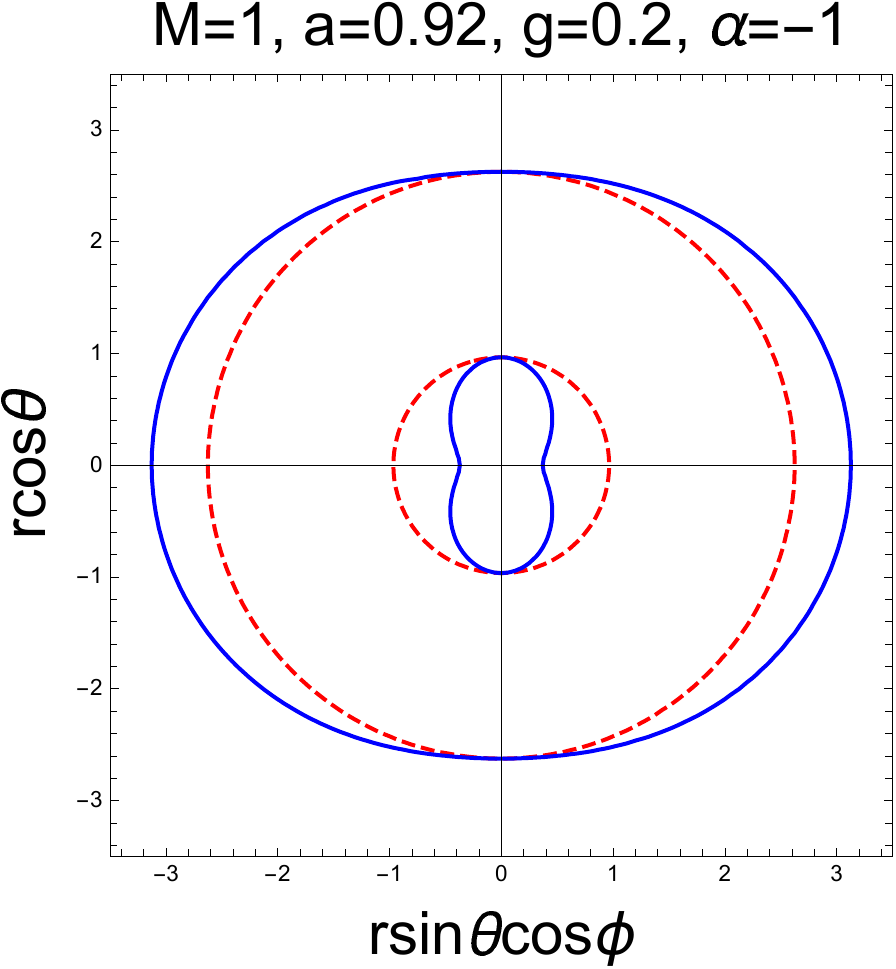}
  		\includegraphics[width=0.24\textwidth]{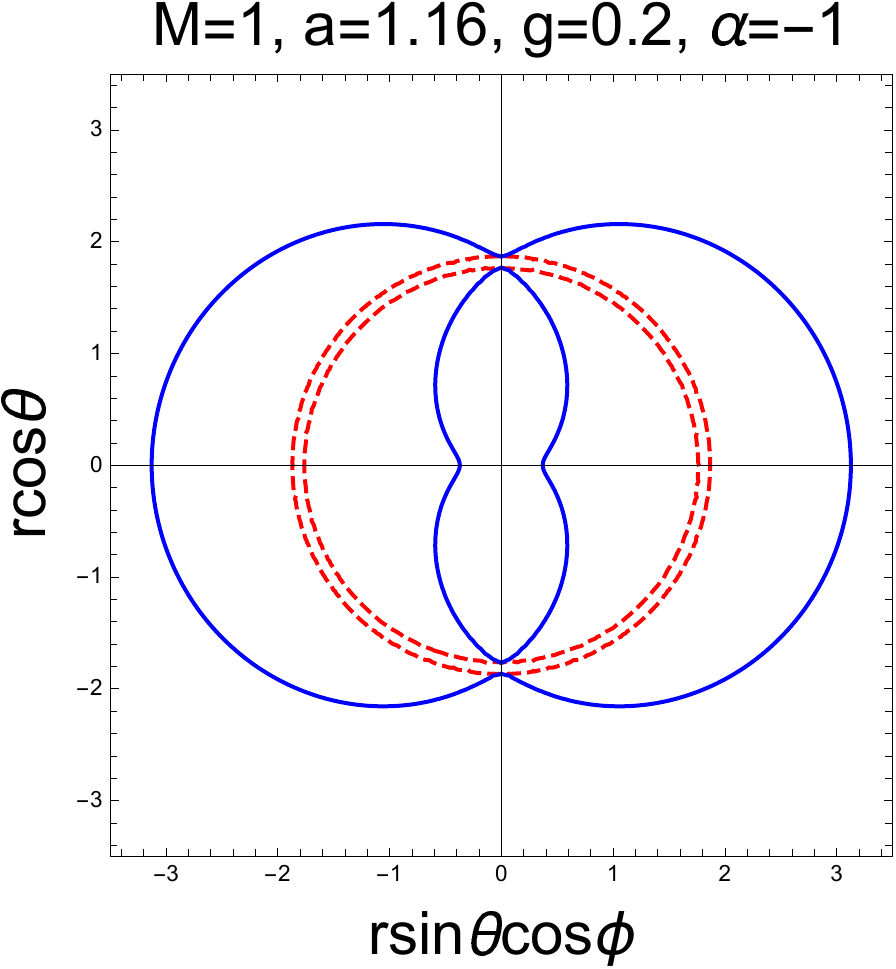}
  		\includegraphics[width=0.24\textwidth]{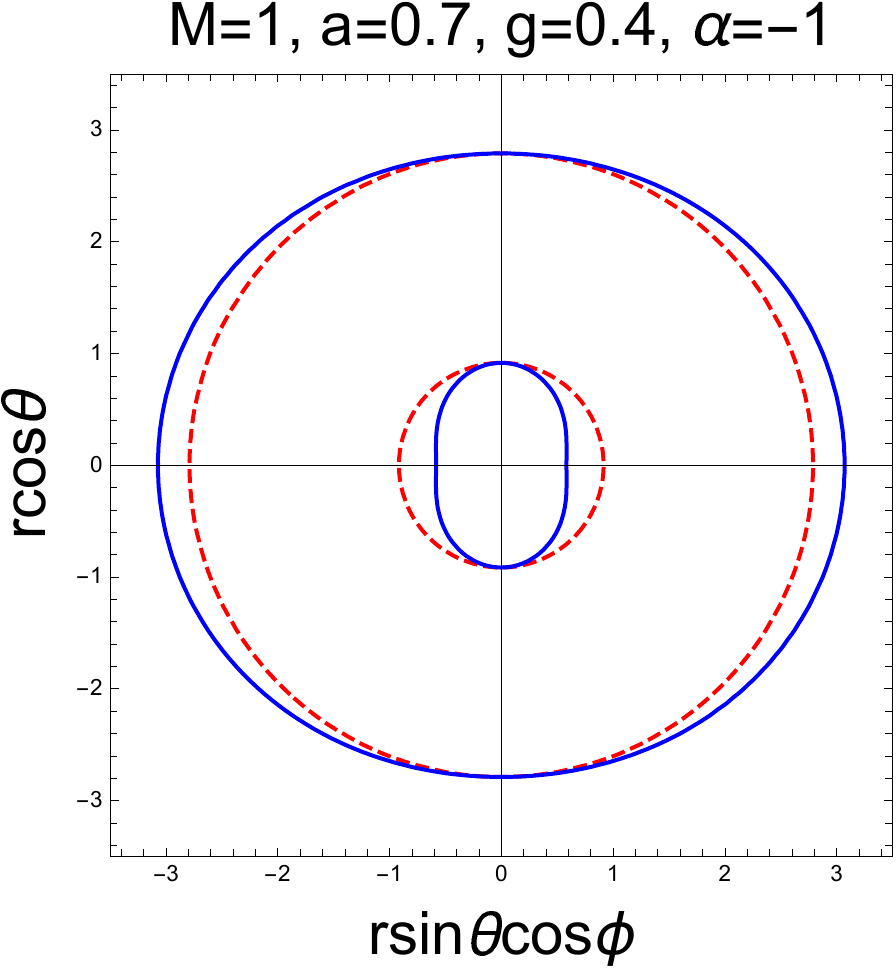}
  		\includegraphics[width=0.24\textwidth]{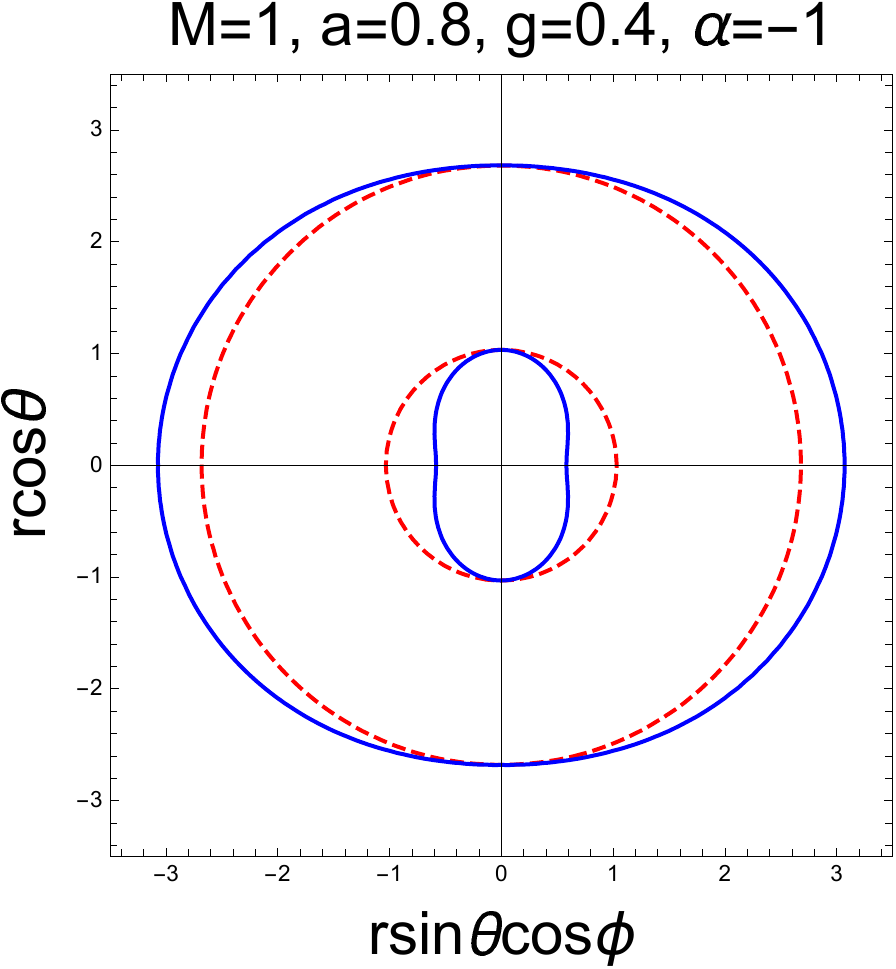}
  		\includegraphics[width=0.24\textwidth]{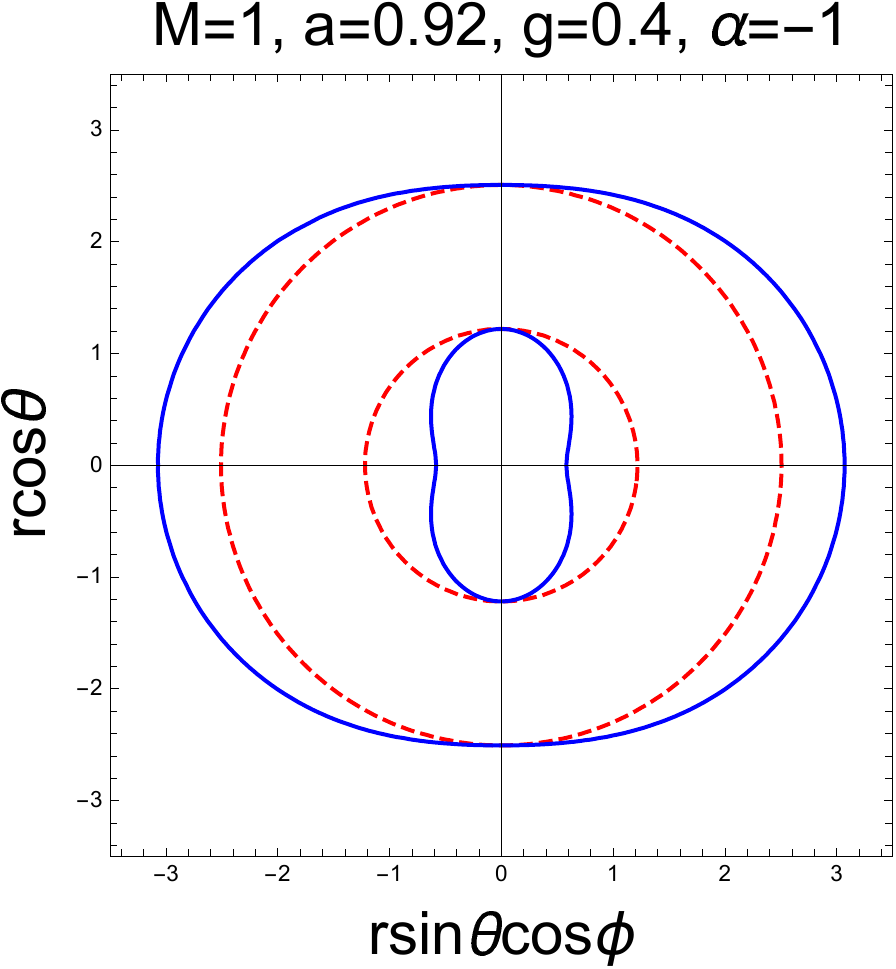}
  		\includegraphics[width=0.24\textwidth]{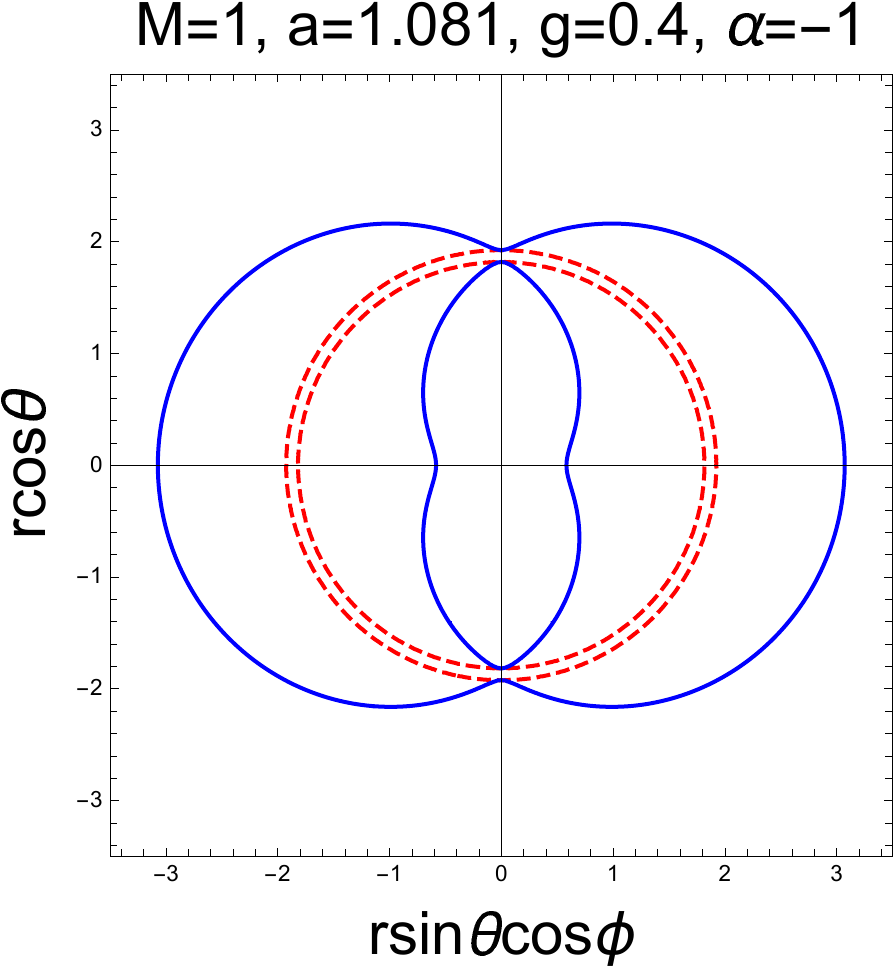}
  \end{figure}

\section{Penrose process}\label{7}
  Since the Killing vector $K^\mu=\partial_t$ becomes space-like inside the ergosphere, there exists the negative energy orbits. Base on this, Penrose first proposed that one can extract the rotational energy from black hole. Consider, for example, a particle $A$ moving in the ergosphere which breaks into two particles $B$ and $C$. We arrange the breakup so that the energy of the particle $B$ falling into the black hole is negative and the particle $C$ escape to infinity. Due to the local conservation of energy along geodesics for this process, the energy of particle $C$ will be greater than that of particle $A$. In order to study Penrose process quantitatively, we have followed Refs.~\cite{haroon2018effects,pradhan2019study,bhattacharya2018kerr}. For simplicity, we will restrict the motion of particles in the equatorial plane ($\theta=\frac{\pi}{2}$).
\par
  The geodesic motion for test particles in the space-time (\ref{eq:rotating metric}) are determined by the following Lagrangian
  \begin{equation}\label{eq:Lagrange}
  2\mathcal{L}=-\left(1-\frac{2\rho}{r}\right)\dot{t}^2-\frac{4a\rho}{r}\dot{t}\dot{\phi}+\frac{r^2}{\Delta_r}\dot{r}^2+\left(r^2+a^2+\frac{2a^2\rho}{r}\right)\dot{\phi}^2,
  \end{equation}
  where $2\rho=\frac{2Mr^{3}}{\left(r^{2}+g^{2}\right)^{\frac{3}{2}}}-\alpha\ln{\frac{r}{\vert\alpha\vert}}$ and $\Delta_{r}=r^{2}+a^{2}-\frac{2Mr^{4}}{\left(r^{2}+g^{2}\right)^{\frac{3}{2}}}+\alpha r\ln{\frac{r}{\vert\alpha\vert}}.$
  In terms of the Euler-Lagrange equation, one gets two conserved quantities, i.e. the energy $E$ and the angular momentum $L$ of the test particle:
  \begin{equation}\label{eq:conserved quantity}
  E\equiv-\frac{\partial\mathcal{L}}{\partial\dot{t}}=-g_{tt}\dot{t}-g_{t\phi}\dot{\phi}~~~\text{and}~~~L\equiv\frac{\partial\mathcal{L}}{\partial\dot{\phi}}=g_{t\phi}\dot{t}+g_{\phi\phi}\dot{\phi},
  \end{equation}
  where the dot ($\cdot$) denotes derivatives with respect to affine parameter $\lambda$.
\par
  The radial equation for the geodesic motion of a test particle can be obtained by means of Eqs.~(\ref{eq:Lagrange}) and (\ref{eq:conserved quantity}) as
  \begin{equation}
  r^4\dot{r}^2=E^2\left(r^4+a^4+a^2(2r^2-\Delta_r)\right)+L^2(a^2-\Delta_r)-4aLE\rho r+\delta r^2\Delta_r,
  \end{equation}
  where $\delta=-1,0,1$ correspond to time-like, null and space-like geodesic respectively.
  Suppose that the breakup happened at the turning point of the geodesic, where $\dot{r}=0$, then from the radial equation for equatorial geodesic it follows that
  \begin{equation}\label{eq:E}
  E=\frac{2a\rho rL\pm r\sqrt{\Delta_r}\sqrt{L^2r^2-\delta(r^4+a^4+a^2(2r^2-\Delta_r))}}{r^4+a^4+a^2(2r^2-\Delta_r)},
  \end{equation}
  and
  \begin{equation}\label{eq:L}
  L=\frac{-2a\rho rE\pm r\sqrt{\Delta_r}\sqrt{E^2r^2+\delta(r^2-2\rho r)}}{\Delta_r-a^2}.
  \end{equation}
  Using Eq.~(\ref{eq:E}), one could derive the condition while the value of the energy is negative: in order to have positive energy in the limit $a\rightarrow0$, we must retain only the positive sign; on the other hand when $a\ne0$, a necessary criterion for having negative energy is $L<0$. Thus in order to have negative energy, we must also have
  \begin{equation*}
  4a^2\rho^2 r^2L^2>r^2\Delta_r\left[L^2r^2-\delta(r^4+a^4+a^2(2r^2-\Delta_r))\right],
  \end{equation*}
  or alternatively
  \begin{equation}
  1-\frac{2\rho}{r}<\delta\frac{\Delta_r}{L^2}\ .
  \end{equation}
  Since $g_{tt}(\theta=\frac{\pi}{2})=-(1-\frac{2\rho}{r})$, the above inequality clearly suggests that this can happen in the ergosphere.
\par
\begin{figure}[htbp]
	\centering
	\includegraphics[width=.8\textwidth]{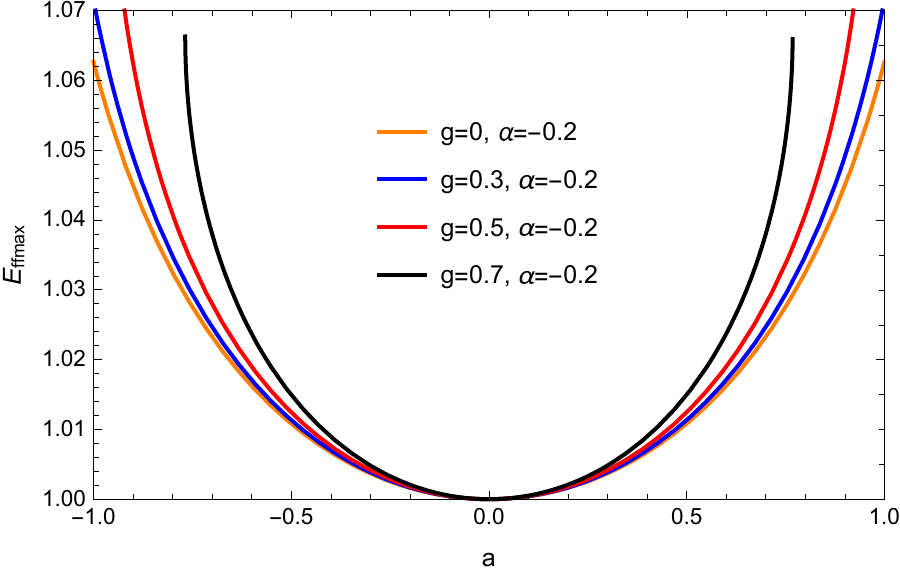}
	\includegraphics[width=.49\textwidth]{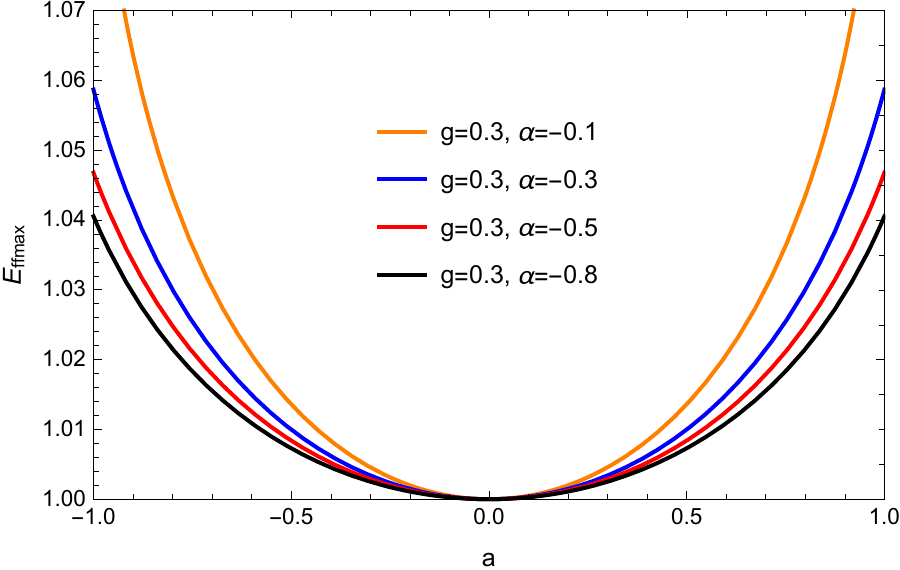}
	\includegraphics[width=.49\textwidth]{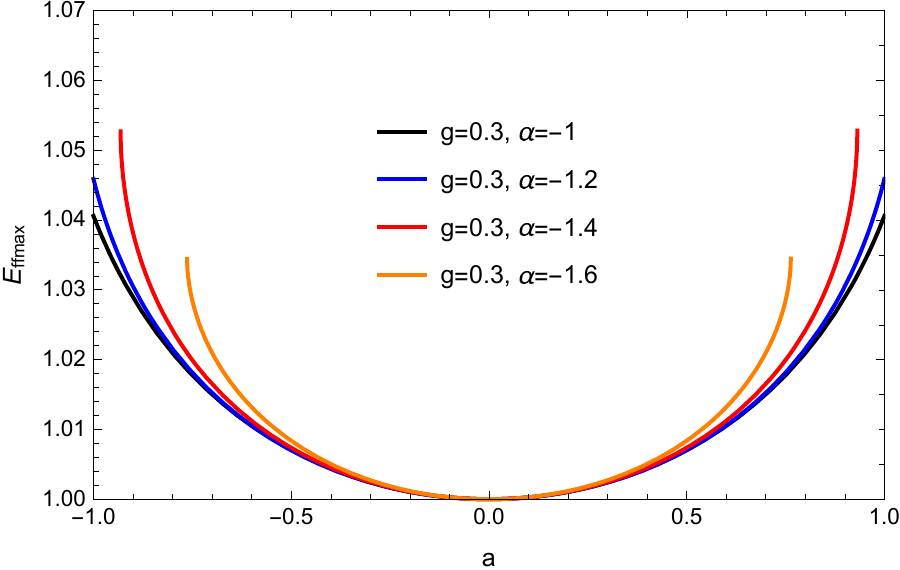}
	\caption{Variation of maximal efficiency of Penrose process with $a$,~$g$ and $\alpha$.}
	\label{energy gain}
\end{figure}

  Let us now consider a massive particle ($\delta=-1$) with $E_A=1>0$ (without loss of generality) and angular momentum $L_A$ entering the ergosphere. This particle then decays into two photons ($\delta=0$) with energies and angular momenta ($E_B,L_B$) and ($E_C,L_C$) respectively. We could arrange this process such that the photon which falls into the event horizon has negative energy and the photon which escapes to infinity has positive energy. From Eq.~(\ref{eq:L}), we have
  \begin{equation}\label{eq:LABC}
  \begin{split}
  L_A&=\frac{-2a\rho r+r\sqrt{\Delta_r}\sqrt{2\rho r}}{\Delta_r-a^2}=\alpha_A,\\
  L_B&=\frac{-2a\rho r-r^2\sqrt{\Delta_r}}{\Delta_r-a^2}E_B=\alpha_B(r) E_B,\\
  L_C&=\frac{-2a\rho r+r^2\sqrt{\Delta_r}}{\Delta_r-a^2}E_C=\alpha_C(r) E_C.
  \end{split}
  \end{equation}
  The conservation of energy and angular momentum gives
  \begin{equation}\label{eq:LB}
  E_B+E_C=E_A=1
  \end{equation}
  and
  \begin{equation}\label{eq:LC}
  L_B+L_C=\alpha_B(r)E_B+\alpha_C(r)E_C=L_A=\alpha_A(r).
  \end{equation}
  Solving Eqs.~(\ref{eq:LB}) and (\ref{eq:LC}) and using Eq.~(\ref{eq:LABC}), one finally gets
  \begin{align}
  E_B&=\frac{1}{2}\left(1-\sqrt{1+\frac{a^2-\Delta_r}{r^2}}\right)=\frac{1}{2}\left(1-\sqrt{\frac{2Mr^{2}}{\left(r^{2}+g^{2}\right)^{\frac{3}{2}}}-\frac{\alpha}{r}\ln{\frac{r}{\vert\alpha\vert}}}\right),\label{eq:EB}\\
  E_C&=\frac{1}{2}\left(1+\sqrt{1+\frac{a^2-\Delta_r}{r^2}}\right)=\frac{1}{2}\left(1+\sqrt{\frac{2Mr^{2}}{\left(r^{2}+g^{2}\right)^{\frac{3}{2}}}-\frac{\alpha}{r}\ln{\frac{r}{\vert\alpha\vert}}}\right).
  \end{align}
  Clearly, the photon $C$ that escapes from the black hole to infinity has more energy than the initial particle $A$, and the energy gain $\Delta E$ in this process is
  \begin{equation}
  \Delta E=\frac{1}{2}\left(\sqrt{\frac{2Mr^{2}}{\left(r^{2}+g^{2}\right)^{\frac{3}{2}}}-\frac{\alpha}{r}\ln{\frac{r}{\vert\alpha\vert}}}-1\right)=-E_B.
  \end{equation}
  It follows from Eq.~(\ref{eq:EB}) that the maximum gain in energy occurs at the event horizon, $\Delta_r=0$ and the maximal efficiency of Penrose process is then given by
  \begin{equation}
  E_{ffmax}=\frac{E_A+\Delta E}{E_A}=\frac{1}{2}\left(1+\sqrt{\frac{2Mr_+^{2}}{\left(r_+^{2}+g^{2}\right)^{\frac{3}{2}}}-\frac{\alpha}{r_+}\ln{\frac{r_+}{\vert\alpha\vert}}}\right),
  \end{equation}
  which can be shown visually in Fig.~\ref{energy gain}. It is easy to see that the maximal efficiency $E_{ffmax}$ increases with the increase of spin parameter $a$ and magnetic parameter $g$. Interestingly, the effect of dark matter parameter $\alpha$ on $E_{ffmax}$ is nonmonotonic. When $\alpha>\alpha^C\left(\text{critical value}\right)$, the maximal efficiency $E_{ffmax}$ decrease with its decrease; however, when $\alpha<\alpha^C$, $E_{ffmax}$ increases with its decrease. In fact, this is caused by the influence of $\alpha$ on the event horizon.

\section{Conclusions and discussions}\label{8}
 In this paper, we have obtained the exact solution of the static spherically symmetric Bardeen black hole surrounded by perfect fluid dark matter and then studied the thermodynamic properties. We first derived the thermodynamic first law and the corresponding Smarr formula by treating the magnetic charge $g$ and dark matter parameter $\alpha$ as thermodynamic variables. Furthermore, we discussed the thermodynamic stability of the black hole by means of heat capacity. The result showed that, there exists a critical radius $r_{+}^{C}$, where the heat capacity diverges and the second order phase transition occurs. The critical radius $r_{+}^{C}$ increases with the magnetic charge $g$ and decreases with the dark matter parameter $\alpha$.
\par
 With the Newman-Janis algorithm, we generalized the static spherically symmetric Bardeen black hole surrounded by PFDM to the corresponding rotating solution. According to the components of the energy-momentum tensor, we found that the weak energy condition is violated near the origin of the  rotating Bardeen black holes surrounded by PFDM, which happens for all the rotating regular black holes. Meanwhile, we constrained the dark matter parameter to $\alpha<0$ in terms of the weak energy condition.
\par
 The structure of the black hole horizons was studied in detail. By solving the relevant equation numerically, we found that for any fixed parameters $g$ and $\alpha$, when $a<a_{E}$, the radii of Cauchy horizons increase with the increasing $a$ while the radii of event horizons decrease with $a$. For $a=a_{E}$, we have an extremal black hole with degenerate horizons. If $a>a_{E}$, no black hole will form. Similarly, for any given values of parameters $a$ and $\alpha$, two horizons get closer first with the increase of $g$, then coincide when $g=g_{E}$ and eventually disappear. Furthermore, for a fixed dark matter parameter $\alpha$, we can obtain a critical curve in the parameter space ($a, g$), which separates the black hole region from the no black hole region. We found that the extremal value of the rotation parameter $a_{E}$ decreases with the magnetic charge parameter $g$. By the analysis to ergospheres, we have seen that the size of the ergosphere increases with the rotation parameter $a$ and increases slightly with the increase of $g$. Moreover, when $a>a_{E}$, we have no ergosphere.
\par
 Finally, the energy extraction was discussed by taking into the Penrose process in rotating Bardeen black hole surrounded by PFDM. We have demonstrated that the maximal efficiency $E_{ffmax}$ increases with the increase of spin parameter $a$ and magnetic parameter $g$. However, the effect of dark matter parameter $\alpha$ on $E_{ffmax}$ is nonmonotonic, which is caused by the influence of $\alpha$ on the event horizon.
\par
 Very recently, first image of the M87* black hole was obtained using the sub-millimeter ``Event Horizon Telescope" based on the very-long baseline interferometry \cite{akiyama2019first}. The observation to black hole shadows will be a useful tool for a better understanding of astrophysical black holes and for testing the modified gravity models as well. Hence, as done in Refs.~\cite{stuchlik2019shadow,he2020shadows,zhang2020optical}, we intend to further constrain the relevant black hole parameters by studying the optical properties of the Bardeen black holes surrounded by perfect fluid dark matter.

\section*{Conflicts of Interest}
  The authors declare that there are no conflicts of interest regarding the publication of this paper.

\section*{Acknowledgments}
  We would like to thank the National Natural Science Foundation of China (Grant No.11571342) for supporting us on this work.
  This work makes use of the Black Hole Perturbation Toolkit.

\section*{References}

 \bibliographystyle{unsrt}
 \bibliography{ref}
\end{document}